\documentclass[journal]{IEEEtran}
\ifCLASSINFOpdf
\usepackage{graphicx}
  \graphicspath{{../pdf/}{../jpeg/}}
\DeclareGraphicsExtensions{.pdf,.jpeg,.png,.eps}
\else
\fi
\usepackage[cmex10]{amsmath}
\usepackage{amssymb}
\usepackage{mathtools}

\usepackage{algorithmic}
\usepackage{algorithm}
\newsavebox{\ieeealgbox}

\usepackage[tight,footnotesize]{subfigure}
\usepackage[acronym]{glossaries}
\newacronym{mimo}{MIMO}{multiple-input multiple-output}
\newacronym{v2v}{V2V}{Vehicle-to-vehicle}
\newacronym{gscm}{GSCM}{geometric-based stochastic channel model}
\newacronym{dmc}{DMC}{dense multipath components}
\newacronym{sp}{SP}{specular paths}
\newacronym{ekf}{EKF}{extended Kalman filter}
\newacronym{sage}{SAGE}{space-alternating generalized expectation-maximization}
\newacronym{io}{IO}{interaction objects}
\newacronym{mle}{MLE}{maximum likelihood estimator}
\newacronym{blue}{BLUE}{best linear unbiased estimator}
\newacronym{tx}{TX}{transmitter}
\newacronym{rx}{RX}{receiver}
\newacronym{ddtf}{DDTF}{double directional transfer function}
\newacronym{apdp}{APDP}{average power delay profile}
\newacronym{dod}{DoD}{direction of departure}
\newacronym{doa}{DoA}{direction of arrival}
\newacronym{3d}{3D}{3-dimensional}
\newacronym{4d}{4D}{4-dimensional}
\newacronym{cdf}{CDF}{cumulative distribution function}
\newacronym{hrpe}{HRPE}{High Resolution Parameter Estimation}
\newacronym{mpc}{MPC}{multipath components}
\newacronym{fim}{FIM}{Fisher Information Matrix}
\newacronym{em}{EM}{Expectation Maximization}
\newacronym{tdm}{TDM}{time-division multiplexed}
\newacronym{snr}{SNR}{signal-to-noise ratio}
\newacronym{pdp}{PDP}{power delay profile}
\newacronym{crlb}{CRLB}{Cramer-Rao lower bound}
\newacronym[plural = EADFs, firstplural=effective array distribution functions (EADF)]{eadf}{EADF}{effective aperture distribution function}
\newacronym{rmse}{RMSE}{root mean squared error}
\newacronym{nsl}{NSL}{normalized sidelobe level}
\newacronym{sa}{SA}{simulated annealing}
\newacronym{dser}{DSER}{Doppler shift estimation range}
\newacronym{mse}{MSE}{mean squared error}
\newacronym{uca}{UCA}{uniform circular array}
\newacronym{t2c}{T2C}{truck-to-car}
\newacronym{t2t}{T2T}{truck-to-truck}
\newacronym{ss}{SS}{sequential switching}
\newacronym{nss}{non-SS}{Nonsequential switching}

\newcommand{\ie}{\emph{i.e.}}
\newcommand{\etal}{\emph{et al. }}

\newcommand\wid{0.8}

\newcommand{\RNum}[1]{\uppercase\expandafter{\romannumeral #1\relax}}
\newcommand{\tx}[1]{\text{#1}}
\newcommand{\B}[1]{\textbf{#1}}
\newcommand{\BS}[1]{\boldsymbol{#1}}
\usepackage[utf8]{inputenc}
\usepackage{color}

\usepackage{amsthm}
\theoremstyle{definition}
\newtheorem{Property}{Property}

\usepackage{siunitx}
\usepackage{booktabs}

\usepackage{soul}
\usepackage{color}

\begin{document}
%
\title{On channel sounding with switched arrays in fast time-varying channels}

\author{\IEEEauthorblockN{Rui Wang, \textit{Student Member, IEEE}, Olivier Renaudin, \textit{Member, IEEE}, \\ C. Umit Bas, \textit{Student Member, IEEE}, Seun Sangodoyin, \textit{Student Member, IEEE}, \\Andreas F. Molisch, \textit{Fellow, IEEE}}}


%


\maketitle

\begin{abstract}
Time-division multiplexed (TDM) channel sounders, in which a single RF chain is connected sequentially via an electronic switch to different elements of an array, are widely used for the measurement of double-directional/MIMO propagation channels. This paper investigates the impact of array switching patterns on the accuracy of parameter estimation of multipath components (MPC) for a time-division multiplexed (TDM) channel sounder. The commonly-used sequential (uniform) switching pattern poses a fundamental limit on the number of antennas that a TDM channel sounder can employ in fast time-varying channels. We thus aim to design improved patterns that relax these constraints. To characterize the performance, we introduce a novel spatio-temporal ambiguity function, which can handle the non-idealities of real-word arrays. We formulate the sequence design problem as an optimization problem and propose an algorithm based on simulated annealing to obtain the optimal sequence. As a result we can extend the estimation range of Doppler shifts by eliminating ambiguities in parameter estimation. We show through Monte Carlo simulations that the root mean square errors of both direction of departure and Doppler are reduced significantly with the new switching sequence. Results are also verified with actual vehicle-to-vehicle (V2V) channel measurements. 

\end{abstract}


%
\IEEEpeerreviewmaketitle

\section{Introduction}
Realistic radio channel models are essential for development and improvement of communication transceivers and protocols \cite{molisch2012wireless}. Realistic models in turn rely on accurate channel measurements. Various important wireless systems are operating in fast-varying channels, such as mobile millimeter wave (mmWave) \cite{HurMagzine}, \gls{v2v} \cite{mecklenbrauker2011vehicular} and high speed railway systems \cite{wang2016channel}. The need to capture the fast time variations of such channels creates new challenges for the measurement hardware as well as the signal processing techniques.

Since most modern wireless systems use multiple antennas, channel measurements also have to be done with \gls{mimo} channel sounders, also known as double-directional channel sounders. There are three types of implementation of such sounders: (i) full \gls{mimo}, where each antenna element is connected to a different RF chain \cite{kim2015large}, (ii) virtual array, where a single antenna is moved mechanically to emulate the presence of multiple antennas \cite{martin1998spatio}, and (iii) switched array, also known as \gls{tdm} sounding, where different physical antenna elements are connected via an electronic switch to a single RF chain \cite{thoma2000identification}. The last configuration is the most popular in particular for outdoor measurements, as it offers the best compromise between cost and measurement duration.

From measurements of MIMO impulse responses or transfer functions, it is possible to obtain the parameters (\gls{dod}, \gls{doa}, delay, and complex amplitude) of the \glspl{mpc} by means of \gls{hrpe} algorithms. Most \gls{hrpe} algorithms are based on the assumption that the duration of one \gls{mimo} snapshot (the measurement of impulse responses from every transmit to every receive antenna element) is shorter than the coherence time of the channel. Equivalently, this means the \gls{mimo} cycle rate, i.e. the inverse of duration between two adjacent \gls{mimo} snapshots, should be greater than or equal to half of the maximal absolute Doppler shift, in order to avoid ambiguities in estimating Doppler shifts of \glspl{mpc}. Since in a switched sounder the \gls{mimo} snapshot duration increases with the number of antenna elements, there seems to be an inherent conflict between the desire for high accuracy of the DoA and DoD estimates (which demands a larger number of antenna elements) and the admissible maximum Doppler frequency.

Yin \etal were the first to realize that the \gls{dser} can be extended by a factor equal to the product of \gls{tx} and \gls{rx} antennas in \gls{tdm} channel sounding \cite{yin2003doppler}. They studied the problem in the context of the ISI-SAGE algorithm \cite{fleury2003high}. Pedersen \etal later discovered that it was the choice of \gls{ss} patterns that caused this limit in the \gls{tdm} channel sounding \cite{pedersen2004joint}. \gls{nss} patterns can potentially significantly extend the \gls{dser} by eliminating the ambiguities. However the analysis is performed under ideal conditions: firstly the paper assumes that all antenna elements are isotropic radiators, secondly it requires the knowledge of the phase centers of all antennas. Both assumptions are difficult to fulfill for realistic arrays used in channel sounding, given the unavoidable mutual coupling between antennas and the presence of a metallic support frame. Pedersen \etal proposed to use the so-called \gls{nsl} of the objective function as the metric to evaluate switching patterns, and further derived a necessary and sufficient condition of the switching sequences that lead to ambiguities \cite{pedersen2008optimization}, but the method to derive good switching sequences for realistic arrays is not clear. Our work attempts to close these gaps.

Our work adopts an algebraic system model that uses decompositions through \glspl{eadf} \cite{belloni2007doa}, which provides a reliable and elegant approach for signal processing on real-world arrays. Therefore our analysis no longer requires the isotropic radiation pattern or prior knowledge about the antenna phase centers. Based on the Type-\RNum{1} ambiguity function for an arbitrary array \cite{eric1998ambiguity}, we propose a spatio-temporal ambiguity function and investigate its properties and impact on the estimation of directions and Doppler shifts of \glspl{mpc}. We also demonstrate that this spatio-temporal ambiguity function is closely related to the correlation function induced from the \gls{mle} utilized in the state-of-the-art RiMAX algorithm. Inspired by Ref. \cite{chen2008mimo}, we model the array pattern design problem as an optimization problem and propose an annealing algorithm to search for an acceptable solution. Besides we propose an \gls{hrpe} algorithm that adopts a signal data model that incorporates the optimized array switching pattern.\footnote{The conference version of this work is accepted and will be presented at IEEE International Conference on Communications 2018 \cite{wang2018sw}.} 

The main contributions of this work are the following:
\begin{itemize}
 \item we model the selection of array switching scheme as an optimization problem, and introduce the spatio-temporal ambiguity function that also incorporates \textit{realistic} arrays with the aid of \glspl{eadf};
 \item we integrate the optimized array switching pattern into an \gls{hrpe} algorithm and compare the parameter estimation variance with the \gls{crlb} through Monte-Carlo simulations;
 \item we also modify the switching pattern on a real-time \gls{mimo} channel sounder and use actual \gls{v2v} measurement data to show the effectiveness of the optimized switching pattern.
\end{itemize}

The remainder of the paper is organized as follows. Section \RNum{2} introduces the general spatio-temporal ambiguity function and the associated signal data model used in the \gls{tdm}-based channel sounding. In Section \RNum{3} we simplify the problem by only allowing the \gls{tx} to have cycle-dependent switching patterns, and present the formulation of the optimization problem and its solution based on a \gls{sa} algorithm. In Section \RNum{4} we introduce an \gls{hrpe} algorithm that generalizes the evaluation of \gls{tdm} MIMO channel measurement and incorporates the optimized switching pattern. Section \RNum{5} validates the switching sequence and corresponding \gls{hrpe} algorithm via extensive Monte Carlo simulations and measured \gls{v2v} channel responses. In Section \RNum{6} we draw the conclusions.

The symbol notation used in this paper follows the rules below.
\begin{itemize}
 \item Bold upper case letters, such as $\B{B}$, denote matrices. Bold lower case letters, such as $\B{b}$, denote column vectors. For instance $\B{b}_j$ is the $j$-th column of the matrix $\B{B}$.
 \item Calligraphic upper-case letters denote high-dimensional tensors.
 \item $[\B{B}]_{ij}$ denotes the element in the $i$th row and $j$th column of the matrix $\B{B}$.
 \item Superscripts $^\tx{T}$ and $^\dagger$ denote matrix transpose and Hermitian transpose.
 \item The operators $\vert f(\B{x}) \vert$ and $\lVert \B{b} \rVert$ denote the absolute value of a scalar-valued function $f(\B{x})$ and the Euclidean norm of a vector $\B{b}$.
 \item The operators $\otimes$, $\odot$ and $\diamond$ denote Kronecker, Schur-Hadamard, and Khatri-Rao products respectively.
\end{itemize}

\section{Signal Model and Ambiguity Function}
\label{Sec_SystemModel}
 \subsection{Signal Data Model}
This work mainly studies the antenna switching sequence in the \gls{tdm} channel sounding problem. We consider $T$ \gls{mimo} measurement snapshots in one observation window, each with $M_f$ frequency points, $M_R$ receive antennas, and $M_T$ transmit antennas. The adjacent \gls{mimo} snapshots are separated by $T_0$. We assume that all scatterers are placed in the far field of both \gls{tx} and \gls{rx} arrays, which also implies that \glspl{mpc} are modeled as plane waves. Besides \gls{tx} and \gls{rx} arrays are vertically polarized by assumption and have frequency-independent responses within the operating bandwidth. Such $T$ \gls{mimo} snapshots can span larger than the coherence time of the channel, but we assume that the structural parameters of the channel, also known as the large-scale parameters, such as path delay, \gls{doa}, \gls{dod} and Doppler shift, remain constant during this period.\footnote{This does not mean that the channel is assumed to be static, instead for each path the phase varies over time due to the presence of Doppler shift.} 

A \textit{vectorized} data model for the observation of $T$ \gls{mimo} snapshots is given in (\ref{Eq:y_VecModel}). It includes contributions from deterministic \gls{sp}s $\B{s}(\BS\theta_{sp})$,  \gls{dmc} $\B{n}_{dmc}$ and measurement noise $\B{n}_0$.
\begin{equation}
 \B{y} = \B{s}(\BS\theta_{sp}) + \B{n}_{dmc} + \B{n}_{0}  \label{Eq:y_VecModel}
\end{equation}
The data model of an observation vector for a number of $P$ \gls{sp}s is determined by
\begin{equation}
 s(\BS\theta_\tx{sp}) = \B{B}(\BS\mu,\BS\eta_T,\BS\eta_R) \cdot \BS\gamma_{vv}, \label{Eq:s_theta}
\end{equation}
where the \gls{sp} parameter vector $\BS\theta_{sp}$ includes the structural parameters $\BS\mu$ and the path weights $\BS\gamma_{vv}$. The basis matrix $\B{B}$ has dimension $M \times P$, and $M$ is the total number of observations given by $M_fM_RM_TT$. The $i$-th column stands for the vectorized basis channel vector for the $i$-th \gls{sp}. The switching sequences $\BS\eta_T$ and $\BS\eta_R$ for \gls{tx} and \gls{rx} respectively not only affect the sequence of samples from different antennas, but also impact the phase variation because of the nature of \gls{tdm} channel sounding. 


The complexity of the basis matrix grows as the irregularity of switching sequence increases. Most papers assume that \gls{ss} sequences are used at both \gls{tx} and \gls{rx}, hence the periodicity can be exploited to greatly simplify $\B{B}$. An example of \gls{ss} sequences is shown in Fig. \ref{fig:SwTiming_Exp}. References \cite{salmi2009detection} and \cite{Wang2015efficiency} adopted a data model of \gls{sp}s where the channel is assumed to be completely static within one \gls{mimo} snapshot, as a result the basis matrix is a Khatri-Rao product of three smaller matrices. To evaluate fast time-varying channel, such as the \gls{v2v} communication channel, a more sophisticated signal model is used in Ref. \cite{WangHRPE2016}, which considers the phase variation due to Doppler shifts between switched antennas, but again the model is only applicable for \gls{ss} sequences. 


\subsection{Spatio-temporal Ambiguity Function}
\label{sect:st_AmbFunc}
The Type-\RNum{1} ambiguity function for an antenna array can reflect its ability to differentiate signals in the angular domain \cite{eric1998ambiguity}. Generalizing the definition to include the full structural parameters $\BS\mu$ from Eq. (\ref{Eq:s_theta}), we have a spatio-temporal ambiguity function  
 \begin{equation}
    X_{tot}(\BS{\mu},\BS{\mu}^\prime) =  \frac{ \B{b}^\dagger(\BS{\mu})\B{b}(\BS{\mu}^\prime)}{\lVert\B{b}(\BS{\mu}) \rVert \cdot \lVert\B{b}(\BS{\mu}^\prime) \rVert}, \label{Eq:Amb_func}
 \end{equation}
where $\B{b}$ is one column of the basis matrix $\B{B}$ in Eq. (\ref{Eq:s_theta}). The ambiguity function also depends on the switching sequences $\BS\eta_T$ and $\BS\eta_R$, but we drop them from the notation for brevity.

More importantly this ambiguity function is closely related with the correlation function that is tightly connected with the objective function of the \gls{mle} developed in Section \ref{subsect:ParamInitialization}, hence studying the properties of this ambiguity function is critical for the performance of the \gls{mle}.

Here we introduce some properties of the ambiguity function.
\theoremstyle{definition}
\begin{Property}
  $X_{tot}(\BS\mu,\BS\mu) = 1$, and $0 \le \big\vert X_{tot}(\BS\mu,\BS\mu^\prime) \big\vert \le 1$. 
  \label{Prop:AmbRange}
\end{Property}
It is straightforward to prove that $X_{tot}(\BS\mu,\BS\mu) = 1$. We can use the Cauchy-Schwartz inequality to prove the inequality $\big\vert X_{tot}(\BS\mu,\BS\mu^\prime) \big\vert \le 1$, and the equality is obtained when $\BS\mu = \BS\mu^\prime$.
\theoremstyle{definition}
\begin{Property}[Separability of the Ambiguity Function]
We can prove that the ambiguity function in Eq. (\ref{Eq:Amb_func}) is a product of two component ambiguity functions of delay $\tau$ and $\BS\kappa$. The latter consists of every parameter in $\BS\mu$ except $\tau$. 
\begin{equation}
 X_{tot}(\BS{\mu},\BS{\mu}^\prime) = X_\tau(\tau,\tau^\prime) X(\BS\kappa,\BS\kappa^\prime) \label{Eq:Amb_Separate}
\end{equation}
The proof relies on the following property of Khatri-Rao products between two vectors. In fact, the Khatri-Rao product between two vectors is equivalent to the Kronecker product.  
\begin{equation}
 (\B{a}\diamond \B{b})^\dagger \cdot (\B{a}^\prime \diamond \B{b}^\prime) = (\B{a}^\dagger \B{a}^\prime) \cdot (\B{b}^\dagger \B{b}^\prime) \label{Eq:Khatri-Rao_Separate}
\end{equation}
This property holds when $\B{a}$ and $\B{a}^\prime$ have the same length, so do $\B{b}$ and $\B{b}^\prime$. Detailed derivations are provided in Appendix \ref{append:Amb_Separability}.
\label{Prop:AmbProd}
\end{Property}
\theoremstyle{definition}
\begin{Property}
The estimation problem has ambiguities when the following condition holds.
\begin{equation}
 \big \vert X(\BS{\kappa},\BS{\kappa}^\prime) \big \vert = 1, \exists \, \BS\kappa^\prime \neq \BS\kappa
\end{equation}
\end{Property}
 
This spatio-temporal array ambiguity function is also closely related to the ambiguity function well studied in \gls{mimo} radar. The \gls{mimo} radar ambiguity function introduced in Ref. \cite{chen2008mimo} allows \gls{tx} to send different waveforms on different antennas, while our problem considers a repeated sounding waveform for all \gls{tx} antennas. The Doppler-(bi)direction ambiguity function introduced in Ref. \cite{pedersen2008optimization} is quite similar to ours. However their ambiguity function assumes that the array has identical elements without any mutual coupling effects, our ambiguity function can handle arbitrary array structures, which suits better for developing \gls{nss} sequences for \gls{mimo} channel sounding.

\subsection{Simplified Signal Data Model}
\label{sect:SimpleSignalModel}
In this subsection we impose some constraints on the switching sequences in order to obtain a more tractable problem. As a comparison the basis matrix $\B{B}$ with the \gls{ss} sequences at both \gls{tx} and \gls{rx} is given by \cite[(20)]{WangHRPE2016}
\begin{equation}
  \B{B}(\BS\mu) = \B{B}_t \diamond \tilde{\B{B}}_{TV} \diamond \tilde{\B{B}}_{RV} \diamond \B{B}_f. \label{Eq:TWC_basisMat}
\end{equation}
One straightforward relaxation on the switching patterns from \gls{ss} sequences is to allow the \gls{tx} array to have a cycle-dependent switching pattern,\footnote{One cycle here means one \gls{mimo} snapshot.} hence the new basis matrix needs to merge the Khatri-Rao product of two basis matrices $\B{B}_t$ and $\tilde{\B{B}}_{TV}$ into $\tilde{\B{B}}_{TV,T}$, in order to reflect such relaxation. 
\begin{align}
  \B{B}(\BS{\mu}) &= \tilde{\B{B}}_{TV,T} \diamond \tilde{\B{B}}_{RV} \diamond \B{B}_f  \label{Eq:Bmatrix_simp} \\
  &=\begin{bmatrix}
  \tilde{\B{B}}_{TV}^1  & \cdots  &\tilde{\B{B}}_{TV}^T 
 \end{bmatrix}^\tx{T}  \diamond \tilde{\B{B}}_{RV} \diamond \B{B}_f,
 \label{Eq:BasisMatrix_Simplify}
\end{align}
where $\tilde{\B{B}}_{TV}^j \in \mathbb{C}^{M_T\times P}$ with $j=1,2,\ldots,T$ represents the spatio-temporal response of the \gls{tx} array at the $j$-th MIMO snapshot, and $\tilde{\B{B}}_{RV} \in \mathbb{C}^{M_R \times P}$ is for the \gls{rx} array with the sequential switching, and $\B{B}_f \in \mathbb{C}^{M_f \times P}$ is the basis matrix that captures the frequency response due to path delay. 

Instead of allowing both \gls{tx} and \gls{rx} array to switch after each sounding waveform (fully-scrambled), we focus on the switching sequences where the \gls{rx} first switches through all possible antennas while the \gls{tx} remains connected with the same antenna. This type of sequences has a relatively simpler basis matrix compared with the fully scrambled case. Compared to (\ref{Eq:Bmatrix_simp}) the fully scrambled case needs to further merge $\tilde{\B{B}}_{TV,T}$ and $\tilde{\B{B}}_{RV}$. Another motivation for studying this type of switching sequences is the efficiency to operate the channel sounder. Many high-power \gls{tx} switches tend to have a longer switching settling time than the \gls{rx} switches. This type of switching sequences that we are interested in effectively limits the number of switching for the \gls{tx} array, reduces the guard time needed in the sounding signal and improves the overall measurement efficiency of the channel sounder.

\begin{figure}[!t]
 \centering
 \includegraphics[width = \wid\columnwidth,clip=true]{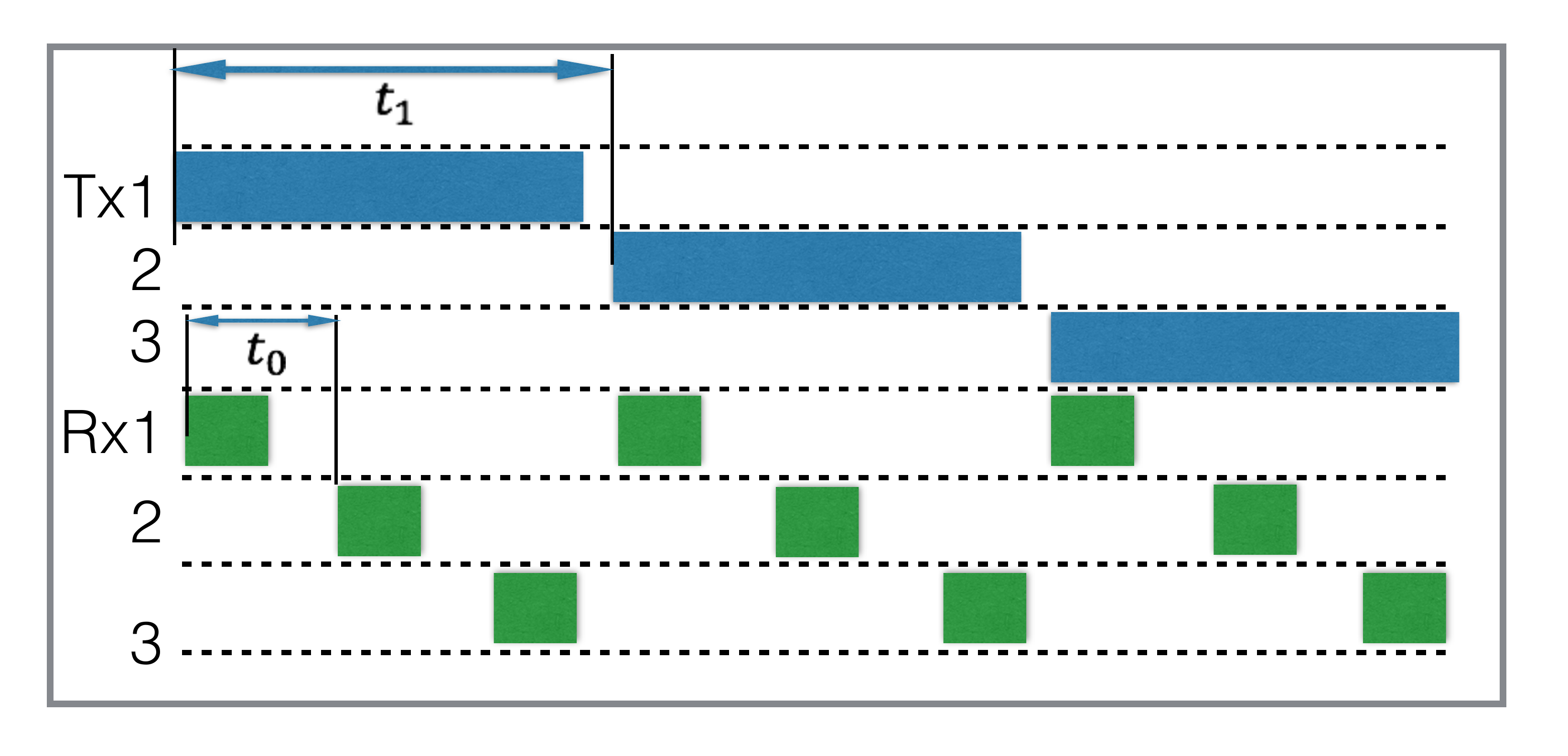}
 \caption{An example of the \gls{ss} switching for the $3 \times 3$ \gls{mimo} setup}
 \label{fig:SwTiming_Exp}
\end{figure}

Since our work focuses on channel sounding in a fast time-varying channel, the phase variation within one \gls{mimo} snapshot is no longer negligible, and the new \gls{tx} or \gls{rx} basis matrices for the $t$-th \gls{mimo} snapshot become weighted versions of the static array basis matrices. The exact connections are given by
\begin{align}
 \tilde{\B{B}}_{TV}^t &= \B{B}_{TV} \odot \B{A}_T^t \\
 \tilde{\B{B}}_{RV} &= \B{B}_{RV} \odot \B{A}_R, 
\end{align}
where $\B{A}_R$ and $\B{A}_T^t$ are weighting matrices that capture the phase change due to effects of Doppler and switching sequences. $\B{A}_T^t$ depends on the \gls{mimo} snapshot index $t$, because \gls{tx} implements a cycle-dependent switching pattern. Let us use a $M_T \times T$ matrix $\BS\eta_T$ to represent the \gls{tx} switching pattern, the elements in the phase weighting matrices $\B{A}_T^t$ and $\B{A}_R$ are given by
\begin{align}
 [\B{A}_T^t]_{m_T,p} &= e^{j2\pi\nu_p [\BS\eta_T]_{m_T,t}} \label{Eq:A_T_t} \\
 [\B{A}_R]_{m_R,p} &= e^{j2\pi\nu_p m_Rt_0},
\end{align}
where $\nu_p$ is the Doppler of the $p$-th path. As shown in the example in Fig. \ref{fig:SwTiming_Exp}, we denote the duration between two switching events as $t_1$ and $t_0$ respectively for the \gls{tx} and \gls{rx} array. The \gls{tx} switching timing matrix $\BS\eta_T$ is given by
\begin{equation}
  [\BS\eta_T]_{m_T,t} = (t-1)M_Tt_1 + ([\B{S}_T]_{m_T,t} - 1)t_1.
\end{equation}
$[\B{S}_T]_{m_T,t}$ takes an integer value between $1$ and $M_T$ and represents the scheduled switching index of the $m_T$-th \gls{tx} antenna for the $t$-th \gls{mimo} snapshot. For the \gls{ss} sequence at \gls{tx}, we have $[\B{S}_T]_{m_T,t} = m_T$, $\forall t=1,2,...,T$. It is not difficult to see that $\tilde{\B{B}}_{TV,T}$ can be broken down into $\B{B}_t\diamond\tilde{\B{B}}_{TV}$ in (\ref{Eq:TWC_basisMat}), which means the basis matrix in (\ref{Eq:BasisMatrix_Simplify}) is a generalized version of (\ref{Eq:TWC_basisMat})

\begin{figure}[!t]
 \centering
 \includegraphics[width = \wid\columnwidth,clip=true]{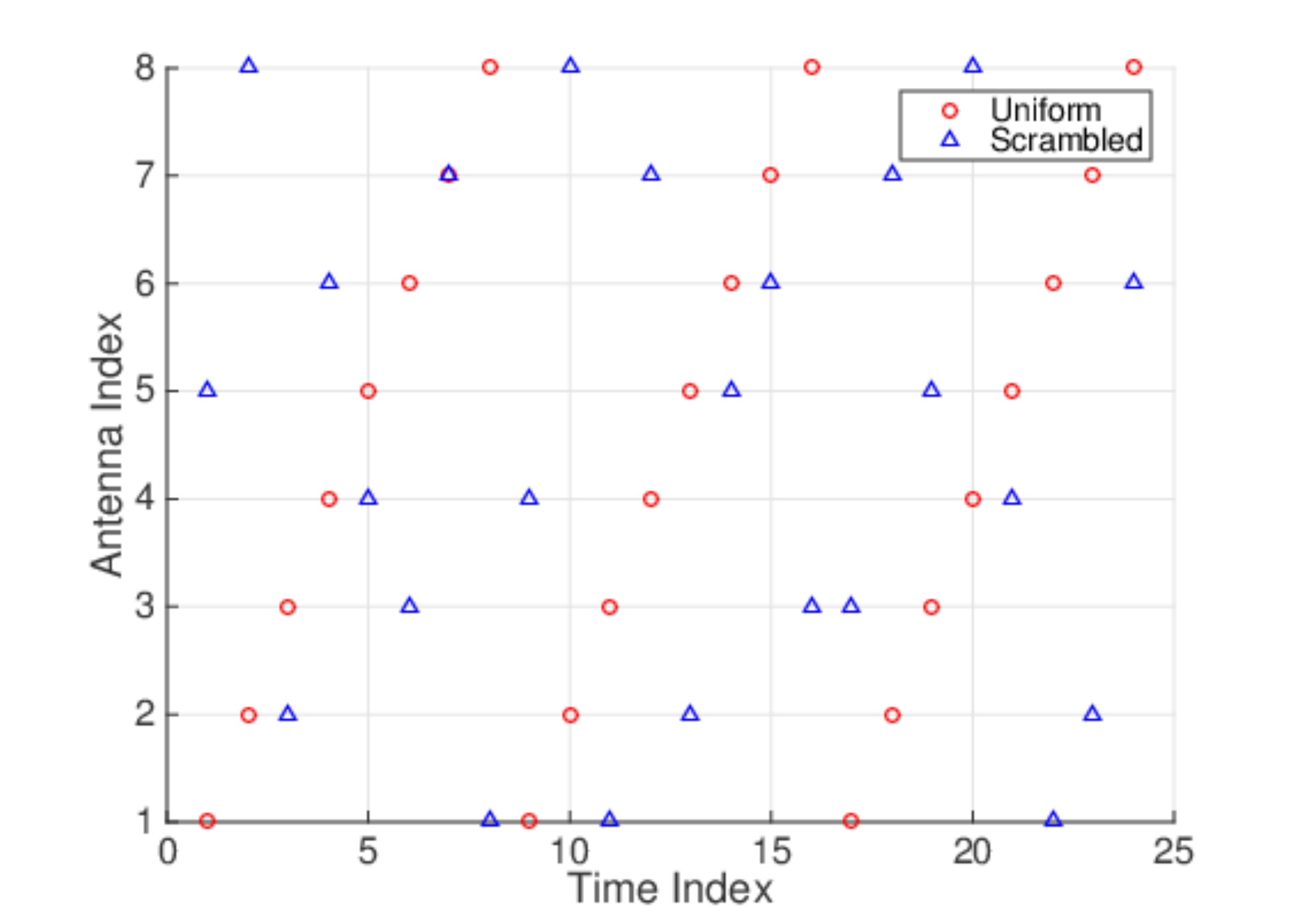}
 \caption{Comparison between the uniform and scrambled \gls{tx} switching sequences with $M_T=8$ and $T=3$}
 \label{fig:TxRS_Exp}
\end{figure}

\section{Switching Pattern Optimization}
\label{Sect:ProblemSetup}
Based on Property 3, it is reasonable to argue that a good switching sequence does not lead to an estimation problem that has ambiguities. Although Ref.  \cite{pedersen2008optimization} focuses on the condition of switching sequences that generate the smallest \gls{crlb}, we have found through simulations that \gls{crlb} is almost identical for the sequential switching and the scrambled switching for practical arrays. Suppressing the sidelobe levels of the ambiguity function within the parameter domain of interest leads to better switching sequences. This is because \gls{crlb} is only relevant for unbiased estimators \cite{kay1993fundamentals}.

Instead of directly evaluating the ambiguity function in (\ref{Eq:Amb_func}), we can find a \textit{upper bound} for its amplitude, which merely depends on the azimuth \gls{dod}, the Doppler Shift, and the \gls{tx} switching pattern. The upper bound is given by
\begin{align}
   \vert X_{tot}(\BS{\mu},\BS{\mu}^\prime) \vert &= \vert X_\tau(\tau,\tau^\prime) \vert \cdot \vert X(\BS\kappa,\BS\kappa^\prime) \vert \\
   & \le \vert X(\BS\kappa,\BS\kappa^\prime) \vert \label{Eq:UpperBound}\\
   & = \vert X_T(\varphi_T,\varphi_T^\prime,\nu,\nu^\prime)X_R(\varphi_R,\varphi_R^\prime, \nu, \nu^\prime) \vert \label{Eq:Amb_TxRxSep}\\
   & \le \vert X_T(\varphi_T,\varphi_T^\prime,\nu,\nu^\prime)\vert, \label{Eq:2ndUpperBound}
\end{align}
where the inequalities in (\ref{Eq:UpperBound}) and (\ref{Eq:2ndUpperBound}) use Property \ref{Prop:AmbRange}, and Eq. (\ref{Eq:Amb_TxRxSep}) uses Property \ref{Prop:AmbProd}. Furthermore we find that the upper bound only depends on the Doppler difference $\Delta\nu = \nu^\prime - \nu$, which is given by 
\begin{align}
 X_T(\varphi_T,\varphi_T^\prime,\nu,\nu^\prime) &= \frac{\tilde{\B{b}}_{TV,T}^\dagger(\varphi_T,\nu) \tilde{\B{b}}_{TV,T}(\varphi_T^\prime,\nu^\prime)}{\lVert \tilde{\B{b}}_{TV,T}^\dagger(\varphi_T,\nu) \rVert \lVert \tilde{\B{b}}_{TV,T}(\varphi_T^\prime,\nu^\prime) \rVert} \label{Eq:AmbFunc_TxNu}\\
 &= X_T(\varphi_T,\varphi_T^\prime,\Delta\nu). \label{Eq:AmbFunc_Sta_nu}
\end{align} 
Appendix \ref{append:Alt_X_T} provides a detailed derivation of Eq. (\ref{Eq:AmbFunc_Sta_nu}), as well as a simpler form of $X_T(\varphi_T,\varphi_T^\prime,\Delta\nu)$.

\begin{figure}[!t]
 \centering
 \includegraphics[width = \wid\columnwidth]{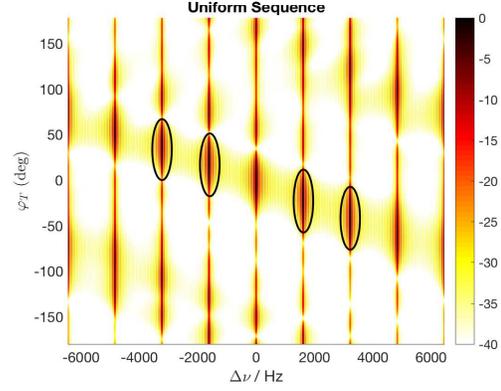}
 \caption{Amplitude of Ambiguity function in dB with Azimuth \gls{dod} and Doppler shift, under uniform switching schemes at both Tx and Rx}
 \label{fig:ChiAmp_unif_exp}
\end{figure}

It is well known that the Doppler shifts and the impinging directions of the plane waves may contribute to phase changes at the output of the array. The Doppler shift leads to a phase rotation at the same antenna when it senses at different time instants, while the propagation direction of the plane wave also contributes to a phase change between two antennas. The \textit{periodic} structure of the uniform switching sequence leads to ambiguities in the joint estimation of Doppler and propagation direction. It is because the estimator may find more than one plausible combination of Doppler and angle that can produce the phase changes over different antennas and time instants. For example, Fig. \ref{fig:ChiAmp_unif_exp} plots the amplitude upper bound given in (\ref{Eq:2ndUpperBound}), when the \gls{tx} uses the uniform switching pattern. We observe multiple peaks in addition to the central peak located at $(0,0)$. It also shows that the non-ambiguous estimation range of Doppler is $[-1/2T_0,1/2T_0)$, when both \gls{tx} and \gls{rx} implement uniform switching patterns and $\varphi_T^\prime = 0$. The value of $T_0$ in this example is $\SI{620}{\mu s}$ which is based on the transmitted signal in Ref. \cite{wang2017real}.

However the design complexity for a fully scrambled switching array may become prohibitive especially when the number of antennas becomes large. Instead we resort to a simplified type of switching sequence introduced in Section \ref{sect:SimpleSignalModel}, where only the \gls{tx} array uses a scrambled switching sequence. 
The \gls{dser} can potentially grow by a factor of $M_T$ from $1/T_0$ to $M_T/T_0$, compared to the maximal boost of $M_TM_R$ in the fully scrambled case. 
Therefore the constraint on the switching sequence simplifies the problem formulation and the corresponding parameter extraction algorithm, while it can still provide a boost on the \gls{dser} that is sufficient for many practical purposes. 

\begin{figure}[!t]
 \centering
 \includegraphics[width = \wid\columnwidth]{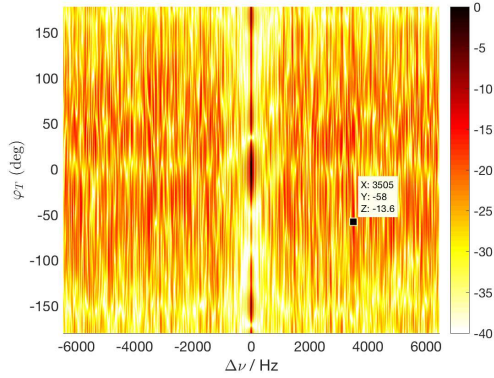}
 \caption{Amplitude of Ambiguity function in dB with Azimuth \gls{dod} and Doppler shift, under uniform Rx switching and scrambled Tx switching scheme}
 \label{fig:ChiAmp_rand_exp}
\end{figure}

\subsection{Problem Formulation}
\begin{figure}[!t] 
 \centering
 \includegraphics[width = \wid\columnwidth]{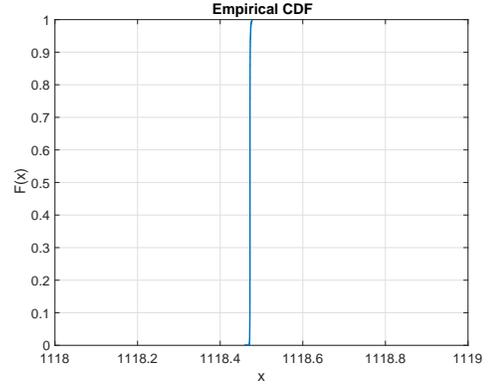}
 \caption{The \gls{cdf} of the energy of the ambiguity function with various \gls{tx} switching schemes $\BS\eta_T$, based on 1000 different configurations}
 \label{fig:CDF_f2}
\end{figure}

The intuitive objective of our array switching design problem is to find schemes that effectively suppress the sidelobes of the spatial-temporal ambiguity function shown in Fig. \ref{fig:ChiAmp_unif_exp}, hence increases the \gls{dser}. Refs. \cite{pedersen2008optimization} and \cite{chen2008mimo} prove that their ambiguity functions have constant energy, so a preferable scheme should spread the volume under the high sidelobes evenly elsewhere. However the proof again uses the idealized assumption about antenna arrays, thus it cannot be applied directly in our case. Here we introduce the function $f_p(\BS\eta_T)$, which is given by 
\begin{align}
 f_p(\BS\eta_T) &= \underset{D}{\iiint} \Big|X_T(\varphi_T,\varphi_T^\prime,\Delta\nu)\Big|^p\; \mathrm{d}\varphi_T  \, \mathrm{d}\varphi_T^\prime  \, \mathrm{d}\Delta\nu, \\
 D &= \{(\varphi_T,\varphi_T^\prime,\Delta\nu) \vert \varphi_T,\varphi_T^\prime \in (-\pi,\pi] \,\& \,\Delta\nu \in [0,\nu_\tx{up}] \}\notag
\end{align}
where $D$ is the integration intervals, and $\nu_\tx{up}$ represents the target maximal Doppler shift below which we want to avoid the ambiguity. This value should equal $M_T/2T_0$, because the periodic sequence is $T_0/M_T$ long for the general cycle-dependent $\BS\eta_T$. Numerical simulations are conducted based on the ambiguity function in Eq. (\ref{Eq:2ndUpperBound}), with the \gls{eadf} of the far-field \gls{tx} array pattern calibrated in the anechoic chamber. The array is one used in a \gls{v2v} \gls{mimo} channel sounder \cite{wang2017real}. Fig. \ref{fig:CDF_f2} suggests that the energy of the ambiguity function, i.e. $f_2(\BS\eta_T)$, is almost constant regardless of the choices of different $\BS\eta_T$. 
As a result we can use $f_p(\BS\eta_T)$ with a higher value of $p$ as the cost function to penalize the \gls{tx} switching schemes that lead to high sidelobes. In summary our optimization problem is given by
\begin{align}
 \underset{\B{S}_T \in \mathcal{C}}{\tx{min}} \; &f_p(\BS\eta_T) \label{Eq:Prob_org}\\
 \text{s.t.}\;  &[\BS\eta_T]_{m_T,t} = ([\B{S}_T]_{m_T,t} - 1)t_1 + (t-1)M_Tt_1, \notag
\end{align}
where the elements in the set $\mathcal{C}$ are integer-valued matrices with a dimension of $M_T \times T$, and every column of $\B{S}_T$ is a permutation of the vector $[1,2,\ldots,M_T]^T$. 

\subsection{Solution and Results}
Because $\B{S}_T$ takes on discrete values in the feasible set, the simulated annealing algorithm is known to solve this type of problem \cite{kirkpatrick1983optimization}. The pseudocode of our proposed algorithm is given here in Alg. \ref{alg:anneal}.
\begin{algorithm}[h]
 \caption{The \gls{sa}-based algorithm to solve the problem (\ref{Eq:Prob_org})}
 \begin{algorithmic}[1]
   \STATE Initialize $\BS\eta_T$, the temperature $T = T_0$, and $\alpha = \alpha_0$;
   \WHILE{ $k\le k_\tx{max}$ or $f_p(\BS\eta_T) > \epsilon_{th}$}
      \STATE $\BS\eta_T^\prime =$ neighbor($\BS\eta_T$);
      \IF{$\tx{exp}\big[ (f_p(\BS\eta_T) - f_p(\BS\eta_T^\prime))/T \big] > \tx{random}(0,1)$}
         \STATE $\BS\eta_T = \BS\eta_T^\prime$
      \ENDIF
      \STATE $T = \alpha T$
   \ENDWHILE
 \end{algorithmic}
 \label{alg:anneal}
\end{algorithm}
The key parameters related to this algorithm are $p=6$, the initial temperature $T_0=100$, the cooling rate $\alpha_0 = 0.97$ and the $k_\tx{max}=500$. The parameters, particularly $\alpha_0$ and $k_\tx{max}$, are selected as a tradeoff between the objective function value and the convergence speed. The operator random(0,1) outputs a random number uniformly distributed between 0 and 1. The operator neighbor($\BS\eta_T$) provides a ``neighbor'' switching pattern by swapping two elements in the same column of $\BS\eta_T$.

The \gls{sa} algorithm implements Markov-Chain Monte Carlo sampling on the discrete feasible set $\mathcal{C}$. If we denote $f_p(\BS\eta_T)$ as $e$ and $f_p(\BS\eta_T^\prime)$ as $e^\prime$. The acceptance probability given by line 4 of Alg. \ref{alg:anneal} is 
\begin{equation}
  P_r(\BS\eta_T,\BS\eta_T^\prime,T) = 
  \begin{cases}
     \tx{exp}\big[ (e-e^\prime)/T \big], &\text{if }e^\prime \ge e \\
     1, & \text{otherwise}
  \end{cases}
\end{equation}

\begin{figure}[!t]
 \centering
 \includegraphics[clip = true, width = \wid\columnwidth]{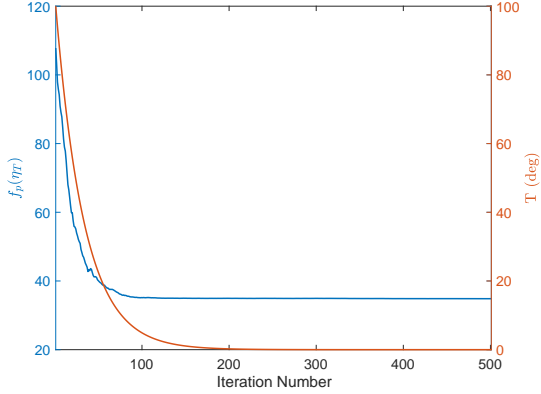}
 \caption{The evolution of $f_6(\BS\eta_T)$ and temperature during the annealing algorithm in Algorithm \ref{alg:anneal}}
 \label{fig:f_p_evolve}
\end{figure}

Fig. \ref{fig:f_p_evolve} provides the values of the objective function with the iteration number and decreased temperature in the \gls{sa} algorithm. We present the amplitude of the 2D ambiguity function (the upper bound given in Eq. (\ref{Eq:2ndUpperBound})) with the final switching sequence in Fig. \ref{fig:ChiAmp_rand_exp}, where the high sidelobes clearly disappear in contrast with Fig. \ref{fig:ChiAmp_unif_exp}. Another useful metric is the \gls{nsl} used in Ref. \cite{pedersen2004joint} to measure the quality of the switching sequence. The \gls{nsl} here is \SI{-13.60}{dB}, while the lowest \gls{nsl} among the three proposed sequences in Ref. \cite{pedersen2004joint} is \SI{-11.06}{dB}. Given the fact that our sequence design approach is not limited to ideal uniform linear arrays and thus more flexible, the reduction of \SI{2.54}{dB} on \gls{nsl} over the best sequence in Ref. \cite{pedersen2004joint} is encouraging.

\section{Parameter Extraction Algorithm}
\label{Sect:HRPE}
This section introduces an \gls{hrpe} algorithm to evaluate the channel sounding data when the \gls{tx} array employs the optimal switching sequence. It is based on the framework of RiMAX \cite{richter2005estimation}, which converges significantly faster than the ISI-SAGE algorithm thanks to the joint optimization of all \glspl{sp}' parameters. For an introduction to the estimation framework we refer to \cite[Sec. \RNum{3}]{WangHRPE2016}. We emphasize the changes made in the parameter initialization algorithm for \gls{sp}s and the local optimization algorithm, when compared to our previous work \cite{WangHRPE2016}. 

\subsection{Path Parameter Initialization}
\label{subsect:ParamInitialization}
The outline of the parameter initialization algorithm of \glspl{sp} is the same with that given in \cite[Alg. 1]{WangHRPE2016}. It is based on the idea of subsequent signal detection, estimation and subtraction. However the main objective function used in signal detection and estimation, also known as the \textit{correlation function}, needs to be adjusted. It is given by
\begin{equation}
     C(\BS{\mu},\B{y}) = (\B{y}^\dagger \B{R}^{\tx{-}1} \B{B}) (\B{B}^\dagger \B{R}^{\tx{-}1} \B{B})^{\tx{-}1} (\B{B}^\dagger \B{R}^{\text{-}1} \B{y}). \label{Eq:Corr_func}
\end{equation}
According to \cite[Alg. 1]{WangHRPE2016}, one important step is to evaluate the correlation function on a $N_f\times N_R \times N_T \times N_t$ multidimensional search grid $\tilde{\BS\mu}_L$. With the increased \gls{dser} $N_t$, the number grid points in the time domain, requires an increase from $2T$ to $2M_TT$. If we compute the correlation function given in (\ref{Eq:Corr_func}) for all possible points in $\tilde{\BS\mu}_L$, we can organize the results into a 4D tensor that shares the same dimension with $\tilde{\BS\mu}_L$. Its $n_t$-th child tensor in the last domain (time domain) can be expressed by 
\begin{equation}
 \mathcal{C}(:,:,:,n_t) = \mathcal{T}_1^{n_t} \odot \mathcal{T}_1^{n_t}  \oslash \mathcal{T}_2^{n_t}, \label{Eq:C_grid_4D}
\end{equation}
where $\oslash$ is the element-wise division between two tensors. Similar to our method in Ref. \cite{Wang2015efficiency}, we can exploit the data structure of $\B{B}$ and $\tilde{\BS\mu}_L$, and apply tensors products to greatly accelerate the computation \cite{TTB_Software}. Appendix \ref{Annex:InitialSPsTensor_RSAA} reveals the detailed procedures on how to compute the two tensors $\mathcal{T}_1^{n_t}$ and $\mathcal{T}_2^{n_t}$.

\subsection{Path Parameter Optimization}
With an estimate of $\B{R}$ and an initial value of $\BS\theta_s$, we further improve $\BS\theta_s$ with the Levenberg-Marquardt (LM) method. The update equation of $\BS\theta_s$ for the $i$-th iteration is given by
\begin{equation}
 \hat{\BS{\theta}}^{i+1}_{s} = \hat{\BS{\theta}}^i_{s} + \big[\mathcal{J}(\hat{\BS{\theta}}^i_{s},\B{R}) + \zeta \B{I} \odot \mathcal{J}(\hat{\BS{\theta}}^i_{s},\B{R})\big]^{\text{-}1} \B{q}(\B{y}|\hat{\BS{\theta}}^i_{s},\B{R}). \label{Eq:LV_Opt_Step}
\end{equation}
This iterative optimization step requires the evaluation of the score function $\B{q}(\B{y}|\BS\theta_s,\B{R})$ and \gls{fim} $\mathcal{J}(\BS{\theta}_s,\B{R})$. Computationally attractive methods are provided in \cite[Sec. \RNum{3}-C]{WangHRPE2016}. The key to apply this method is to rewrite the Jacobian matrix $\B{D}(\BS\theta_s)$ as a sum of Khatri-Rao products and apply the property in (\ref{Eq:Khatri-Rao_Separate}), but an update is needed because of the new signal data model given in (\ref{Eq:s_theta}) and (\ref{Eq:BasisMatrix_Simplify}). Details about the Jacobian matrix can be found in Appendix \ref{Append:JacobMat}. The expression is given by
\begin{equation}
 \B{D}(\BS\theta_\tx{s}) = \frac{1}{2}(\B{D}_3^1 \diamond \B{D}_2^1 \diamond \B{D}_1 + \B{D}_3^2 \diamond \B{D}_2^2 \diamond \B{D}_1), \label{Eq:JacobMat_RSAA}
\end{equation}
where the details of each small $\B{D}$ matrix are summarized in Tab. \ref{Tab:DMatrix}. To reconstruct one small $\B{D}$ matrix from this table, one can concatenate the matrices related to $\B{D}$ along the row direction. Each element in the table has $P$ columns, where $P$ is the number of \glspl{sp}. 

 \begin{table*}[!t]
 \centering
 \renewcommand{\arraystretch}{1.5}
 \caption{Components to compute the Jacobian matrix in (\ref{Eq:JacobMat_RSAA})}
 \label{Tab:DMatrix}
 \begin{tabular}{c|cccccc}
     \toprule
      & $\BS\tau$ & $\BS\varphi_T$ & $\BS\varphi_R$ & $\BS\nu$ & $\BS\gamma_{vv,r}$ & $\BS\gamma_{vv,i}$\\
      \midrule
 $\B{D}_1$ & $\B{D}_f\diamond \BS\gamma_{vv}^T$ & $\B{B}_f \diamond \BS\gamma_{vv}^T$ & $\B{B}_f\diamond \BS\gamma_{vv}^T$ & $\B{B}_f \diamond \BS\gamma_{vv}^T$ & $\B{B}_f $ & $j\B{B}_f $ \\
     $\B{D}_2^1$ & $\tilde{\B{B}}_{RV}$ &  $\tilde{\B{B}}_{RV}$ &  $\B{D}_{\varphi_R,V} \odot \B{A}_R$ & $\tilde{\B{B}}_{RV}$ & $\tilde{\B{B}}_{RV}$ & $\tilde{\B{B}}_{RV}$ \\
      $\B{D}_2^2$ & $\tilde{\B{B}}_{RV}$ &  $\tilde{\B{B}}_{RV}$ &  $\B{D}_{\varphi_R,V} \odot \B{A}_R$ & $\B{B}_{RV} \odot \B{D}_{\nu,R}$ & $\tilde{\B{B}}_{RV}$ & $\tilde{\B{B}}_{RV}$ \\
     $\B{D}_3^1$ & $\tilde{\B{B}}_{TV,T}$ & $\tilde{\B{D}}_{\varphi_T,T}$ & $\tilde{\B{B}}_{TV,T}$ & $2\tilde{\B{D}}_{\nu,T}$ & $\tilde{\B{B}}_{TV,T}$ & $\tilde{\B{B}}_{TV,T}$ \\
     $\B{D}_3^2$ & $\tilde{\B{B}}_{TV,T}$ & $\tilde{\B{D}}_{\varphi_T,T}$ & $\tilde{\B{B}}_{TV,T}$ & $2\tilde{\B{B}}_{TV,T}$ & $\tilde{\B{B}}_{TV,T}$ & $\tilde{\B{B}}_{TV,T}$ \\
     \bottomrule
 \end{tabular}
\end{table*}

\section{Validation}
In this section we use both Monte Carlo simulations and actual \gls{v2v} measurement data to validate the choice of the switching sequence from Section \ref{Sect:ProblemSetup} and the performance of \gls{hrpe} algorithm described in Section \ref{Sect:HRPE}, which we call RiMAX-RS for brevity. We compare it with the \gls{hrpe} algorithm in Ref. \cite{WangHRPE2016} and will use RiMAX-4D to represent it.

\subsection{Simulation}
\begin{figure}[!ht]
  \centering
   \includegraphics[width = \wid\columnwidth,clip=true]{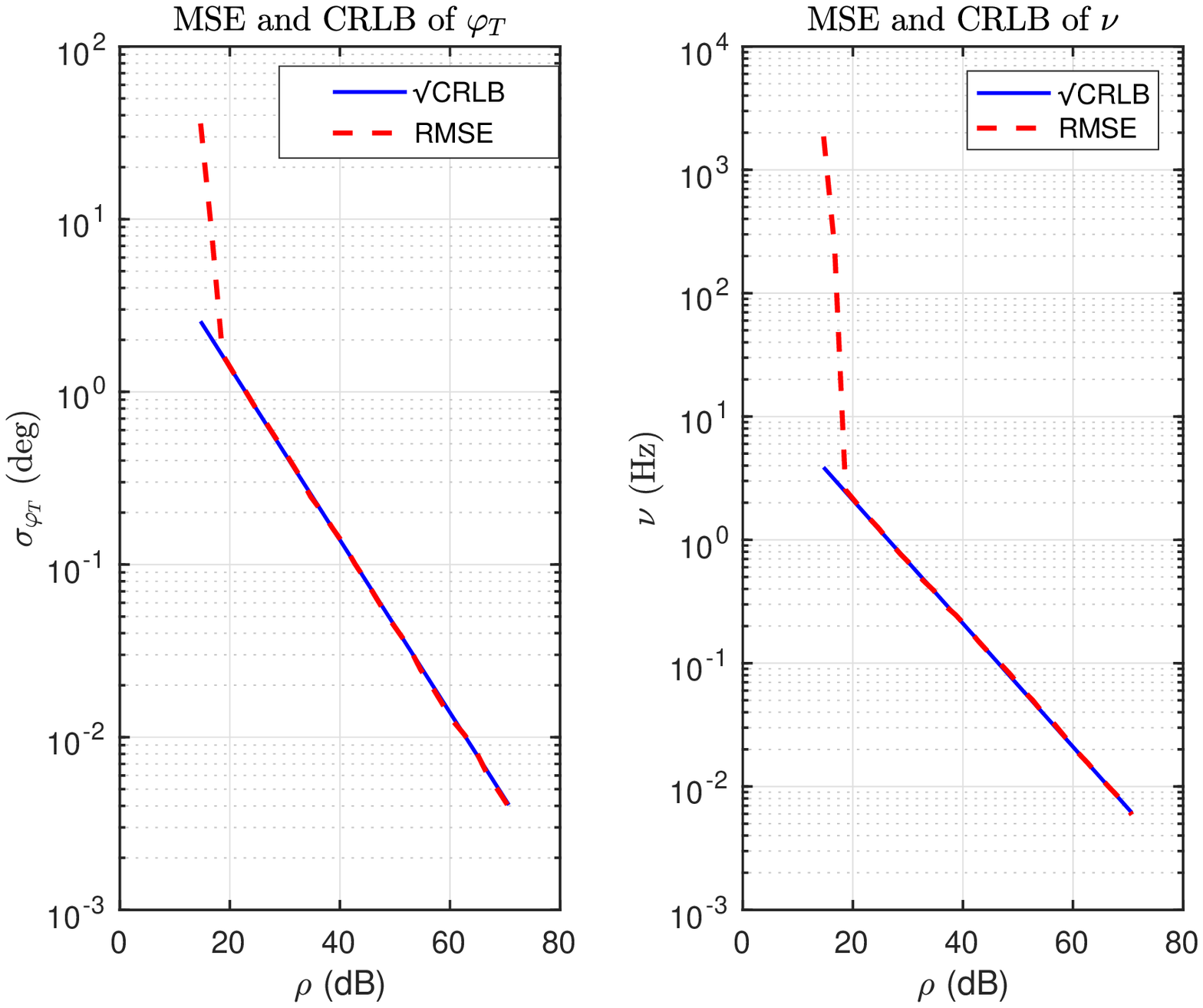}
  \caption{Comparison about \glspl{rmse} of $\varphi_T$ and $\nu$ based on Monte-Carlo simulations with the theoretical \gls{crlb} from (\ref{Eq:crlb_theta_s}) for snapshot 1 in Tab. \ref{Tab:one-path_Ch_param} for the optimized \gls{nss} sequence $\BS\eta_{T,f}$ and RiMAX-RS}
  \label{fig:crlb_case1_rs}
\end{figure}

\begin{figure}[!ht]
  \centering
  \includegraphics[width = \wid\columnwidth,clip=true]{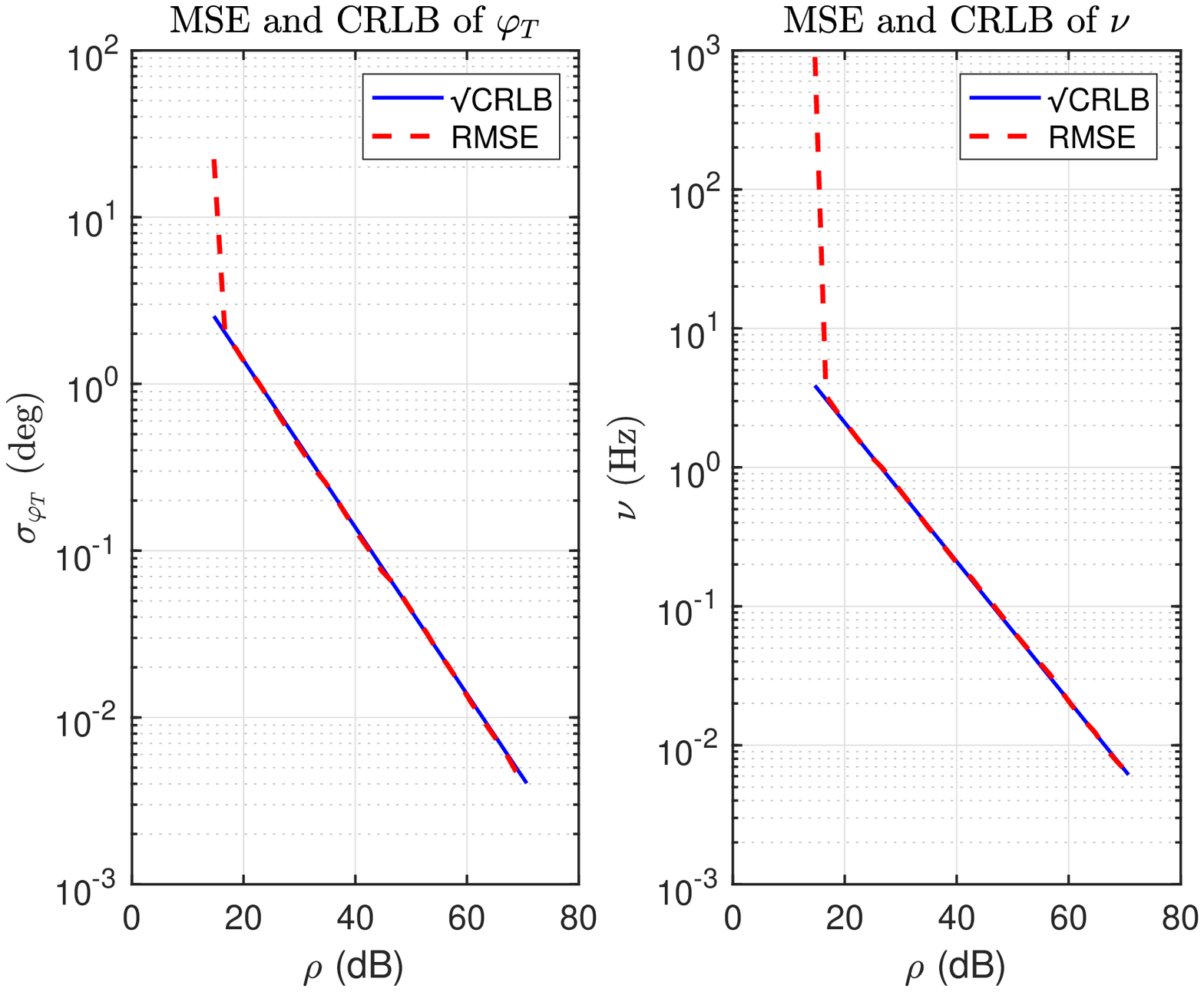}
  \caption{Comparison about \glspl{rmse} of $\varphi_T$ and $\nu$ based on Monte-Carlo simulations with the theoretical \gls{crlb} from (\ref{Eq:crlb_theta_s}) for snapshot 2 in Tab. \ref{Tab:one-path_Ch_param} for the optimized \gls{nss} $\BS\eta_{T,f}$ and RiMAX-RS}
  \label{fig:crlb_case2_rs}
\end{figure}

\begin{figure}[!ht]
  \centering
   \includegraphics[width = \wid\columnwidth,clip=true]{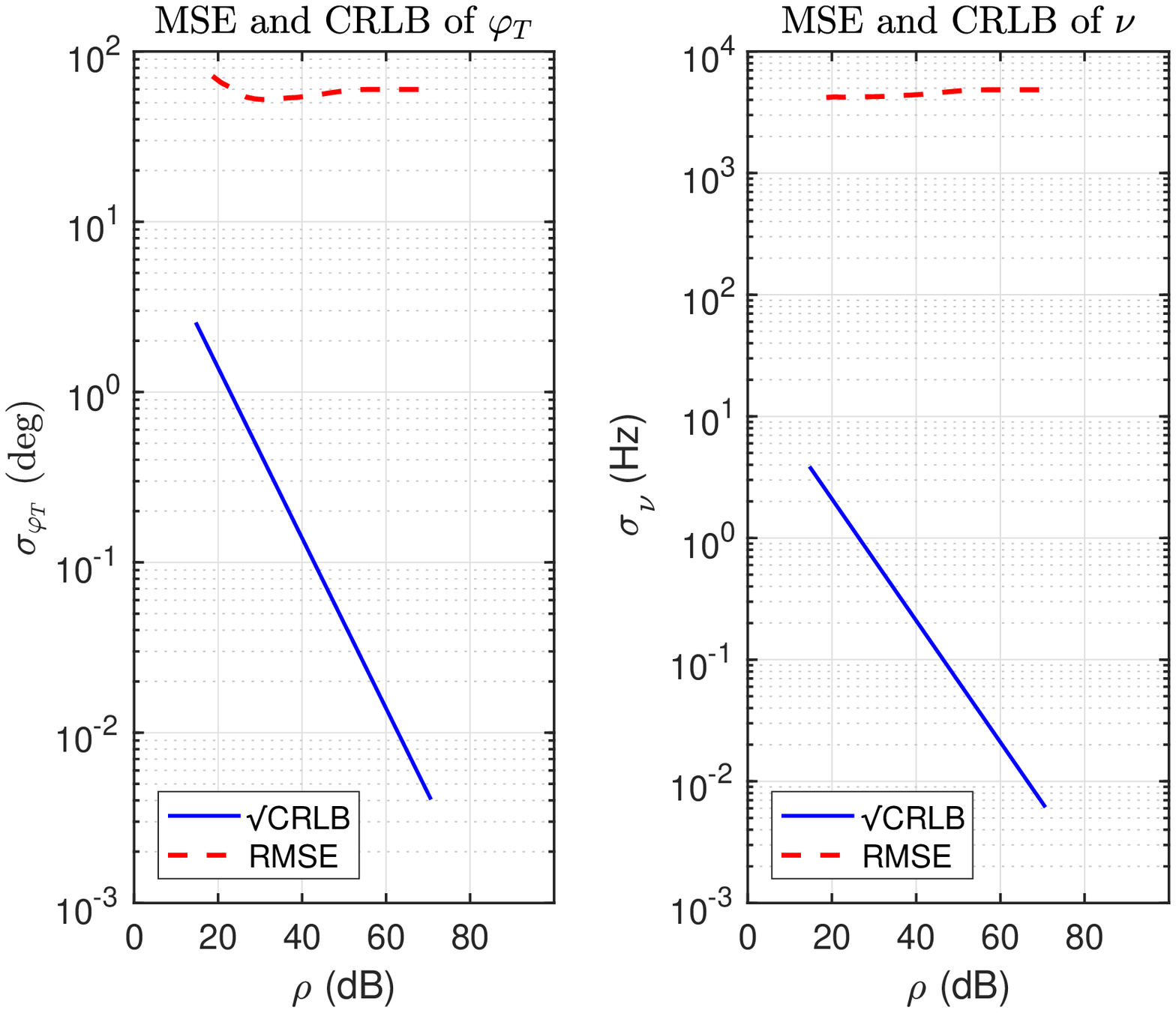}
  \caption{Comparison about \glspl{rmse} of $\varphi_T$ and $\nu$ based on Monte-Carlo simulations with the theoretical \gls{crlb} from (\ref{Eq:crlb_theta_s}) for snapshot 1 in Tab. \ref{Tab:one-path_Ch_param} for the \gls{ss} sequence $\BS\eta_{T,u}$ and RiMAX-4D}
  \label{fig:crlb_case1_unif}
\end{figure}

\begin{figure}[!ht]
  \centering
  \includegraphics[width = \wid\columnwidth,clip=true]{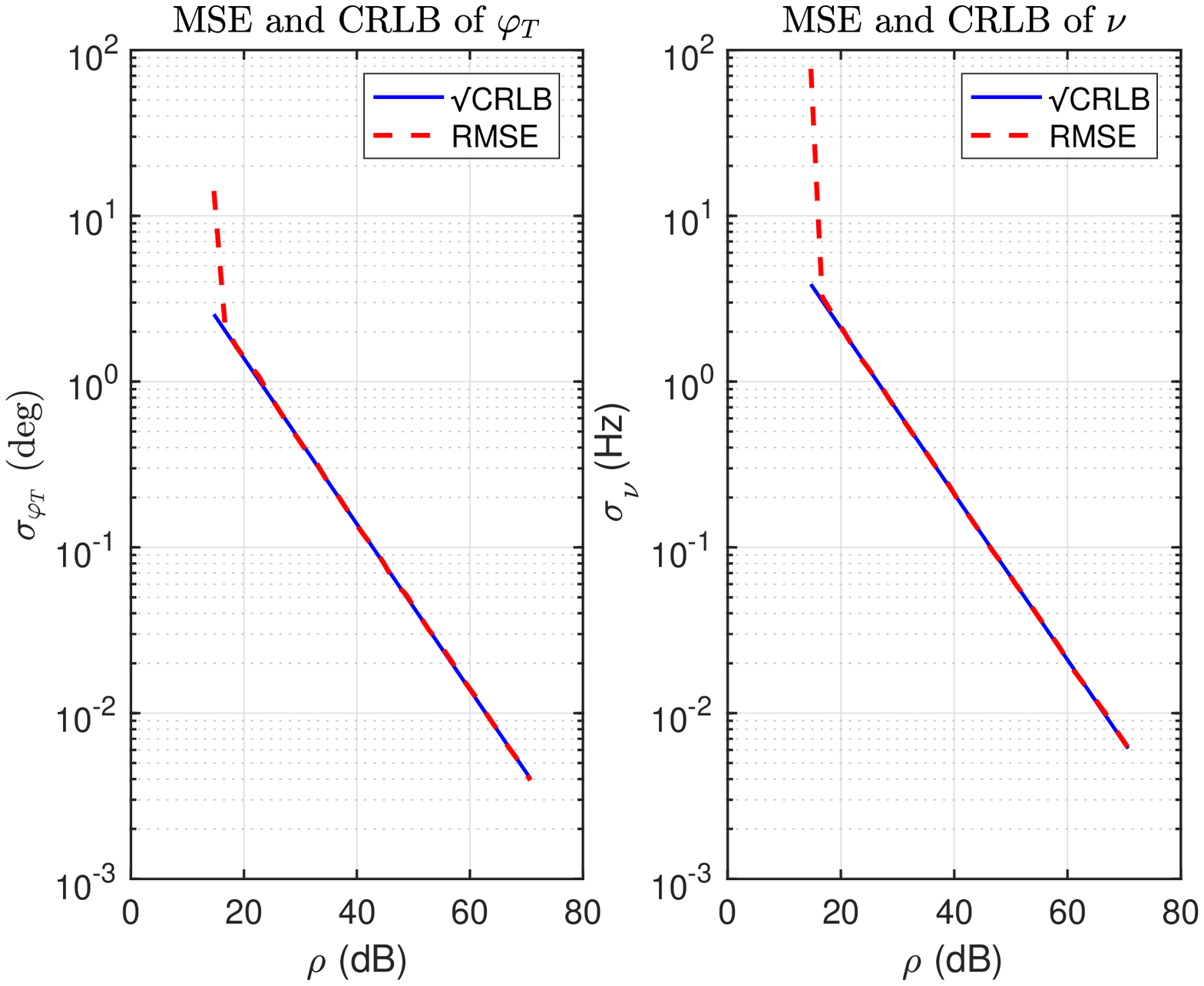}
  \caption{Comparison about \glspl{rmse} of $\varphi_T$ and $\nu$ based on Monte-Carlo simulations with the theoretical \gls{crlb} from Eq. (\ref{Eq:crlb_theta_s}) for snapshot 2 in Tab. \ref{Tab:one-path_Ch_param} for the \gls{ss} sequence $\BS\eta_{T,u}$ and RiMAX-4D}
  \label{fig:crlb_case2_unif}
\end{figure}

First we simulate the single-path channel, whose parameters are listed in Tab. \ref{Tab:one-path_Ch_param}. To cover several cases of interest, the Doppler shift is larger than $1/2T_0\approx\SI{806}{Hz}$ in snapshot 1 and smaller than $1/2T_0$ in snapshot 2, where $T_0$ equals $\SI{620}{\mu s}$ based on the configurations in Ref. \cite{wang2017real}. We compare the \glspl{rmse} with the squared root of the \gls{crlb} as a function of \gls{snr} $\rho$ for two switching sequences, which are the \gls{ss} sequence $\BS{\eta}_{T,u}$ and our optimized \gls{nss} \gls{tx} sequence $\BS{\eta}_{T,f}$. The \gls{snr} is evaluated according to 
\begin{equation}
  \rho = \frac{\lVert \B{s}(\BS\theta_{s}) \rVert^2}{\sigma_n^2}  
\end{equation} 
At each \gls{snr} value, the path weight $\gamma$ is scaled accordingly and 1000 realizations of the channel are generated and subsequently estimated with RiMAX-RS. The theoretical \gls{crlb} can be determined based on \gls{fim} and given by
\begin{equation}
  \sigma_{\BS\theta_s}^2 \succeq \tx{diag}(\mathcal{J}^\tx{-1}(\BS\theta_s)),
  \label{Eq:crlb_theta_s}
\end{equation}
where $\succeq$ is the generalized inequality for vectors. Figs. \ref{fig:crlb_case1_rs} and \ref{fig:crlb_case2_rs} provide such a comparison for $\BS{\eta}_{T,f}$, which demonstrates its good performance in both channels with high or low Doppler. On the other hand, Fig. \ref{fig:crlb_case1_unif} shows the poor estimation accuracy in the high Doppler case for $\BS{\eta}_{T,u}$, although the \gls{mse} can achieve the \gls{crlb} in the low Doppler scenario as expected in Fig. \ref{fig:crlb_case2_unif}. 

\begin{figure}[!ht]
  \centering
  \includegraphics[width = \wid\columnwidth,clip=true]{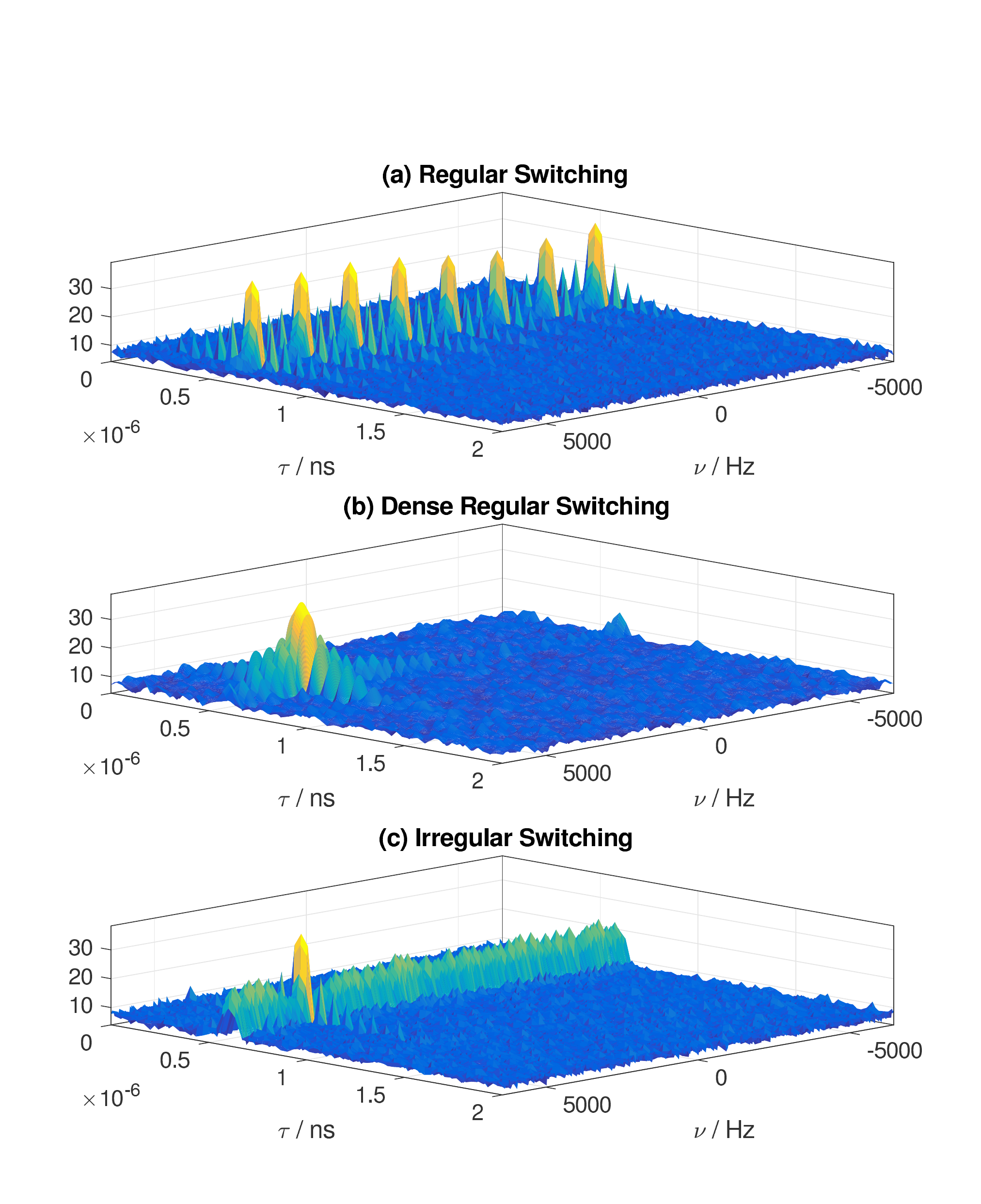}
  \caption{Delay-Doppler spectrum of snapshot 1 with the large Doppler under different \gls{tx} switching sequences, (a) $\BS\eta_{T,u}$ (b) $\BS\eta_{T,d}$ (c) $\BS\eta_{T,f}$}
  \label{fig:Delay_Doppler_Spec}
\end{figure}

We also show in Fig. \ref{fig:Delay_Doppler_Spec} the delay-Doppler spectrum of snapshot 1 with three different \gls{tx} switching sequences, which are $\BS\eta_{T,u}$, $\BS\eta_{T,d}$ (known as the ``dense'' sequential sequence with \gls{mimo} snapshot duration reduced to $T_0/8$), and $\BS\eta_{T,f}$. As a result, the spectrum in Fig. \ref{fig:Delay_Doppler_Spec}(a) displays multiple peaks in the same delay bin but at different Doppler shifts, while Fig. \ref{fig:Delay_Doppler_Spec}(c) shows that $\BS\eta_{T,f}$ successfully eliminates all the peaks except for one at the desired location. Notice that $\BS\eta_{T,f}$ also helps distribute the power under those unwanted peaks equally across Doppler. Fig. \ref{fig:Delay_Doppler_Spec}(b) shows that with $\BS\eta_{T,d}$, we can also eliminate the repeated main peaks (and achieve slightly lower sidelobe energy); however, at the price of the separation time between adjacent \gls{mimo} snapshots is reduced to $T_0/8$ in $\BS\eta_{T,u}$, which would imply that the number of antenna elements would have to be reduced such that $M_T M_R$ decreases by a factor of 8.

\begin{table}[!t]
  \centering
  \caption{Path parameters of one-path channel with high Doppler in snapshot 1 and small Doppler in snapshot 2}
  \label{Tab:one-path_Ch_param}
  \begin{tabular}{l|c|c|c|c}
    \toprule
    Snapshot & $\tau$ (ns) & $\varphi_T$ (deg) & $\varphi_R$ (deg) & $\nu$ (Hz) \\
    \midrule
    1        &  601.1      & 11.5 & 59.6 & 4032.3  \\
    2        &   1117.3    & 21.3 & 160.0 & 80.6   \\
    \bottomrule 
  \end{tabular}
\end{table}

Besides we simulate a two-path channel, which is the simplest version of the multipath channel. Tab. \ref{Tab:case5_est} provides a comparison between the true and estimated parameters where we apply $\BS{\eta}_{T,f}$ and RiMAX-RS. Although both paths' Doppler shifts are larger than $1/2T_0$ with a small difference, the simulation suggests that the estimated parameters are close to the true values.

\begin{table*}
  \small
  \centering
 \caption{Parameter estimation for a two-path channel with $\BS{\eta}_{T,f}$, true/estimate}
 \label{Tab:case5_est}
  \begin{tabular}{c|c|c|c|c|c}
    \toprule
    Path ID & $\tau$ (ns) & $\varphi_T$ (deg) & $\varphi_R$ (deg) & $\nu$ (Hz) & $\vert\gamma\vert^2$ (dB) \\
    \midrule
    1& 646.2/646.2 & 67.81/67.79 & -59.33/-59.33 & 3225.8/3225.8 & -13.13/-13.13\\
    2& 1203.7/1203.7 & -60.15/-60.15 & -123.79/-123.78 & 3217.7/3217.8 & -18.82/-18.82\\
    \bottomrule
  \end{tabular}  
\end{table*}

\subsection{Measurement}

The measurement campaign uses a real-time \gls{mimo} channel sounder developed at the University of Southern California (USC). The sounder is equipped with a pair of software defined radios (National Instruments USRP-RIO) as the main transceivers, two GPS-disciplined rubidium references as the synchronization units and a pair of 8-element \glspl{uca}. The sounding signal is centered at \SI{5.9}{GHz} with a bandwidth of \SI{15}{MHz}. The maximum transmit power is \SI{26}{dBm}. The main advantage of this sounder setup, compared to another \gls{v2v} sounder introduced in Ref. \cite{abbas2011directional}, is the fast \gls{mimo} snapshot repetition rate, which provides a more accurate representation of the channel dynamics though at the price of reduced bandwidth. More details about the sounder setup can be found in Ref. \cite{wang2017real}.

\begin{figure}[!ht]
  \centering
  \subfigure[The front $\&$ side view]{\includegraphics[width = 0.35\columnwidth, 
  clip = true]{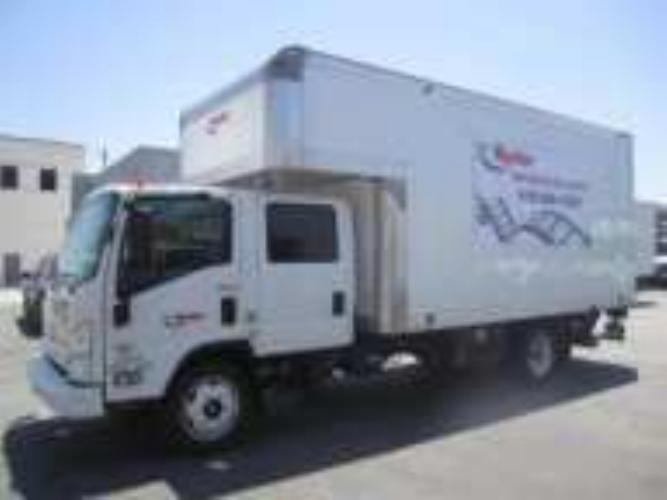}\label{fig:Truck_SideView}}
  \subfigure[Array on top of the cabin]{\includegraphics[width = 0.35\columnwidth, 
  clip = true]{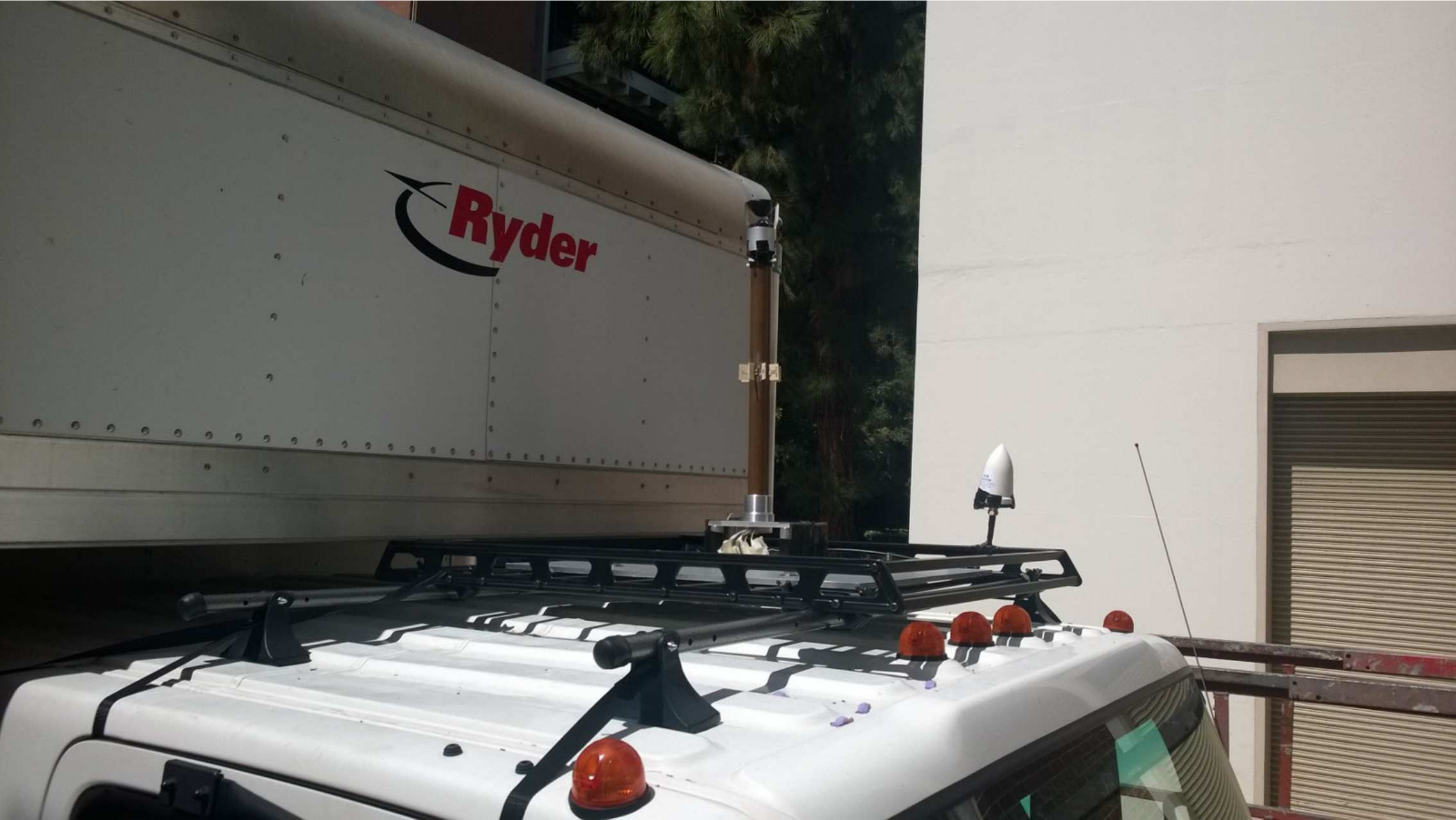}\label{fig:ArrayTopCabin}}
  \caption{The truck used in the \gls{t2c} and \gls{t2t} channel measurements} 
  \label{fig:Truck_pic} 
\end{figure}

We presented first \gls{mimo} measurement results on \gls{t2c} propagation channel at \SI{5.9}{GHz} in Ref. \cite{wang2017pimrc}. The \gls{tx} unit in the channel sounder was programmed to measure with switching sequences alternating between $\BS\eta_{T,u}$ and $\BS\eta_{T,f}$. Therefore the odd \gls{mimo} burst snapshots in the data files were measured with $\BS\eta_{T,u}$, while the even ones were with $\BS\eta_{T,f}$. The adjacent snapshots were \SI{50}{ms} apart, and it is expected that most of large scale parameters remain the same over that timescale. For the truck involved in the \gls{t2c} channel measurement, we use the \SI{5}{m} studio trucks as our test vehicles. Fig. \ref{fig:Truck_pic} shows a picture of the truck and the installation of the array on top of the driving cabin. Each truck has a load capacity about \SI{2722}{kg} and up to $\SI{27}{m^3}$ of cargo space. The sufficient space in the driving cabin allowed us to place the equipment rack of the sounder inside. The platform that holds \gls{tx} or \gls{rx} antenna arrays is tightly clamped on metallic cross-bars installed on top of the driving cabin, in order to ensure the safety of the array and reduce the vibration while we drive the truck, see Fig. \ref{fig:ArrayTopCabin}. 

\begin{figure}[!ht]
  \centering
  \includegraphics[width = \wid\columnwidth, clip = true]{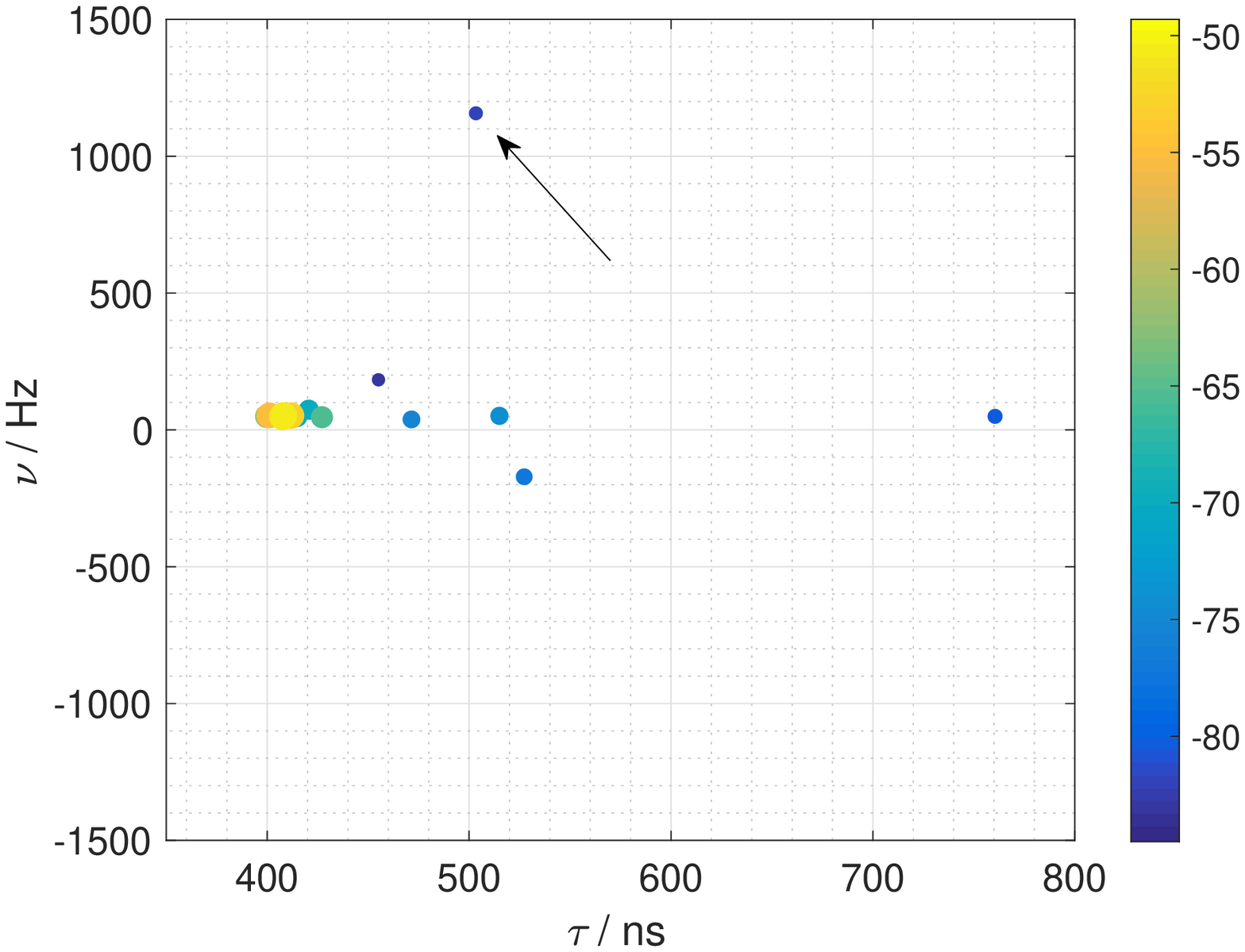}
  \caption{The delay and Doppler spectrum at \SI{896.45}{s} with each point representing a path extracted by RiMAX-RS from Sec. \ref{Sect:HRPE} and color-coded based on the path power} 
  \label{fig:TauNuSpec_RS}
\end{figure}

\begin{figure}[!ht]
  \centering
  \includegraphics[width = \wid\columnwidth, clip = true]{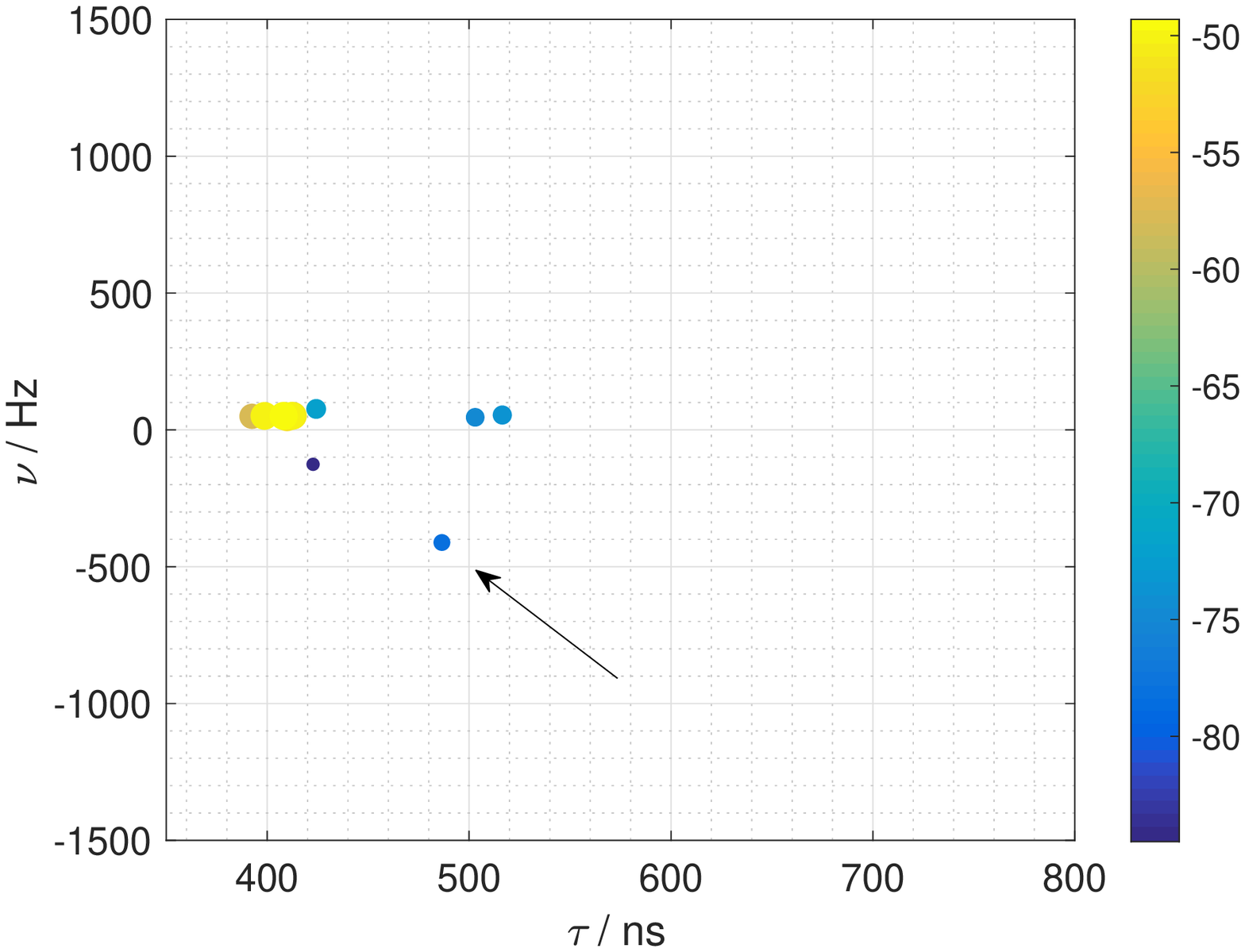}
  \caption{The delay and Doppler spectrum at \SI{896.40}{s} with each point representing a path extracted by RiMAX-4D in Ref. \cite{WangHRPE2016} and color-coded based on the path power}
  \label{fig:TauNuSpec_iid}
\end{figure}

To demonstrate the capability of the optimized switching sequence and the associated \gls{hrpe} algorithm, we analyze several snapshots of the \gls{t2c} channel measurement on I-110 North freeway near Downtown Los Angeles, California, USA. Both the truck (\gls{tx}) and SUV (\gls{rx}) drive in the same direction with an approximate speed of \SI{10}{m/s} estimated based on the recorded GPS locations. An incoming large truck driving in the opposite direction creates a reflected signal path associated with a large Doppler shift. Fig. \ref{fig:TauNuSpec_RS} shows the delay and Doppler spectrum of extracted paths from RiMAX-RS, where we can observe the existence of a weak \gls{mpc} with a large Doppler shift that is outside $[\SI{-806}{Hz},\SI{806}{Hz})$, and it is most likely a reflection from the incoming large truck based on the delay, the large positive Doppler, and angular estimates. The incoming truck can also be observed on a video recording at a time corresponding to the snapshot. As a comparison, Fig. \ref{fig:TauNuSpec_iid} shows the spectrum for a snapshot that is captured \SI{50}{ms} ahead and processed with RiMAX-4D suitable for the \gls{ss} sequence in \cite{WangHRPE2016}. Most of the dominant signals including the line-of-sight path and the reflection from the \gls{tx} truck's trailer are present in both plots, and their Doppler shifts are around \SI{50}{Hz} and delays are between \SI{400}{ns} and \SI{450}{ns}. However there is no component in Fig. \ref{fig:TauNuSpec_iid} to match the \gls{mpc} with large positive Doppler greater than \SI{1}{kHz} in Fig. \ref{fig:TauNuSpec_RS}, instead there exists an \gls{mpc} with a similar delay but a \textit{negative} Doppler around \SI{-416}{Hz}. The difference between two Doppler shifts is about \SI{1568}{Hz}, which is close to $1/T_0\approx \SI{1613}{Hz}$ given by the length of the \gls{dser}. These experimental results also agree with our simulation results in Fig. \ref{fig:Delay_Doppler_Spec} that the conventional \gls{ss} sequence leads to multiple correlation peaks that are $1/T_0$ apart.


\section{Conclusion}
In this paper we revisited the array switching design problem in the context of channel sounding for fast-time varying channels, where the popular \gls{ss} sequences were shown to limit the \gls{dser}. To study the problem in the context of real-world switched arrays, we use \glspl{eadf} for the array modeling which directly incorporates the array nonidealities. We analyzed specifically setups in which the \gls{tx} array implements the \gls{nss} sequence while the \gls{rx} still uses the \gls{ss} sequence. This design constraint reduces significantly the complexity of the \gls{hrpe} algorithm and potentially increases the operation efficiency of the \gls{tdm} channel sounder, while still providing a considerable boost on the \gls{dser}. Future works can possibly look into the trade-off analysis between the operation efficiency and the increased gain on the \gls{dser}. 

To find good non-SS sequences, we formulate the switching sequence design problem as an optimization problem that aims to reduce the high sidelobes in the spatio-temporal ambiguity function, and the proposed \gls{sa} algorithm with carefully selected parameters can provide a \gls{nss} sequence with a sufficiently low \gls{nsl}. 
We integrate the switching sequence into a state-of-the-art parameter extraction algorithm. Both the switching sequence and the corresponding \gls{hrpe} algorithm are verified through extensive Monte-Carlo simulations and actual sample data from \gls{t2c} channel measurements on highway. This shows the capability of our design method to facilitate \gls{v2v} \gls{mimo} measurements with more antennas and mm-wave \gls{mimo} measurements in dynamic environments.

\appendices
\section{Derivation of Separability of the Ambiguity Function} 
\label{append:Amb_Separability}
If we start with the basis vector based on the basis matrix in Eq. (\ref{Eq:Bmatrix_simp}) and apply the property in Eq. (\ref{Eq:Khatri-Rao_Separate}) and, we can further express the inner product between two basis vectors by
\begin{align}
 \B{b}^\dagger(\BS\mu) \B{b}(\BS\mu^\prime) = & \big[\tilde{\B{b}}_{TV,T}^\dagger (\Omega_T,\nu) \tilde{\B{b}}_{TV,T} (\Omega_T^\prime, \nu^\prime) \big] \cdot \notag \\
 & \big[\tilde{\B{b}}_{RV}^\dagger(\Omega_R,\nu) \tilde{\B{b}}_{RV}(\Omega_R^\prime,\nu^\prime) \big] \cdot \notag \\   & \big[\B{b}_f^\dagger (\tau) \B{b}_f(\tau^\prime)\big], \label{Eq:AmbFunc_Nominator_Sep}
\end{align}
where we see that the inner product is factored into three inner products of small basis vectors from different data domains, $\ie$ frequency ($f$), \gls{rx} array ($RV$), and \gls{tx}-plus-time ($TV,T$). 

Similarly we can also express the denominator of the ambiguity function in Eq. (\ref{Eq:Amb_func}) with vector inner products,
\begin{equation}
 \lVert \B{b}(\BS\mu) \rVert \cdot \lVert \B{b}(\BS\mu^\prime) \rVert = \sqrt{\big[\B{b}^\dagger(\BS\mu) \B{b}(\BS\mu)\big] \cdot \big[\B{b}^\dagger(\BS\mu^\prime) \B{b}(\BS\mu^\prime)\big] },
\end{equation}
then we can apply the separability feature derived from Eq. (\ref{Eq:AmbFunc_Nominator_Sep}) with $\BS\mu^\prime=\BS\mu$ and obtain the following
\begin{align}
 \lVert \B{b}(\BS\mu) \rVert \cdot \lVert \B{b}(\BS\mu^\prime) \rVert = &\lVert \tilde{\B{b}}_{TV,T}(\Omega_T,\nu) \rVert \cdot \lVert \tilde{\B{b}}_{RV}(\Omega_R,\nu) \rVert \cdot \notag \\
 &\lVert \tilde{\B{b}}_{TV,T}(\Omega_T^\prime,\nu^\prime) \rVert \cdot \lVert \tilde{\B{b}}_{RV}(\Omega_R^\prime,\nu^\prime) \rVert \cdot \notag \\
 &\lVert \B{b}_f(\tau) \rVert \cdot \lVert \B{b}_f(\tau^\prime) \rVert \label{Eq:AmbFunc_Denominator_Sep}
\end{align} 

From Eqs. (\ref{Eq:AmbFunc_Nominator_Sep}) and (\ref{Eq:AmbFunc_Denominator_Sep}) we show that both the numerator and denominator in Eq. (\ref{Eq:Amb_func}) can be factored into a product of two parts. The first one is only associated with delay $\tau$, and the other contains the rest of parameters in $\mu$ except $\tau$. This means we prove that the ambiguity function is a product of two component ambiguity functions, which is given in Eq. (\ref{Eq:Amb_Separate}).

\section{A Simpler Expression of $X_T$}
\label{append:Alt_X_T}
In this appendix we provide a simpler expression of the essential ambiguity function $X_T(\varphi_T,\varphi_T^\prime,\nu,\nu^\prime)$ given by Eq. (\ref{Eq:AmbFunc_Sta_nu}). This simplification is made possible by exploiting the structure of $\tilde{\B{b}}_{TV,T}$, when \gls{tx} implements a cycle-dependent switching pattern and \gls{rx} uses a sequential switching pattern.

The derivation of Eq. (\ref{Eq:AmbFunc_Sta_nu}) requires another property related with the Hadamard-schur product of two column vectors. It is given by
\begin{equation}
  (\B{a}_1 \odot \B{b}_1)^\dagger (\B{a}_2 \odot \B{b}_2) = (\B{a}_1\odot \B{a}_2^*)^\dagger (\B{b}_1^* \odot \B{b}_2),
\end{equation}
where $\B{a}_1$, $\B{a}_2$, $\B{b}_1$ and $\B{b}_2$ have the same length. Using this property we can express the numerator of Eq. (\ref{Eq:AmbFunc_TxNu}) by
\begin{align}
 &\sum_{t = 1}^T \tilde{\B{b}}_{TV,T}^t (\varphi_T,\nu) ^\dagger \tilde{\B{b}}_{TV,T}^t(\varphi_T^\prime,\nu^\prime) \notag \\
 &=  \sum_{t=1}^T [\B{b}_{TV}(\varphi_T) \odot e^{j2\pi\nu\BS\eta_T^t}]^\dagger [\B{b}_{TV}(\varphi_T^\prime) \odot e^{j2\pi\nu^\prime \BS\eta_T^t}] \notag \\
 &= \Big[\B{b}_{TV}(\varphi_T) \odot \B{b}_{TV}^*(\varphi_T^\prime)\Big]^\dagger \sum_{t=1}^T e^{j2\pi(\nu^\prime - \nu)\BS\eta_T^t}
\end{align}

Besides the Euclidean norm in the denominator of Eq. (\ref{Eq:AmbFunc_TxNu}) can be efficiently evaluated by substituting $\varphi_T^\prime$ with $\varphi$ and $\nu^\prime$ with $\nu$.
\begin{align}
 \lVert \tilde{\B{b}}_{TV,T} (\varphi_T,\nu) \rVert &= \sqrt{\tilde{\B{b}}_{TV,T}(\varphi_T,\nu)^\dagger \tilde{\B{b}}_{TV,T} (\varphi_T,\nu)} \notag \\
 &= \sqrt{T} \cdot \lVert \B{b}_{TV}(\varphi_T) \rVert
\end{align}

\section{Ambiguity function and Correlation function}
\label{append:amb_corre_func}
This appendix provides the relationship between the ambiguity function $X_{tot}$ and the correlation function $C(\BS\mu,\B{y})$ in Eq. (\ref{Eq:Corr_func}). Assuming there is one \gls{sp}, we replace $\B{B}$ with $\B{b}$ in the derivation. Furthermore with large \gls{snr} the observation vector \B{y} can be replaced by its mean or $\B{b}(\BS\mu^\prime)\gamma_{vv}$. Finally we also assume that $\B{R} = \sigma_n^2 \B{I}$. After incorporating three assumptions we have
\begin{align}
 C(\BS\mu,\B{y}(\BS\mu^\prime)) &= \frac{1}{\sigma_n^2}(\B{y}^\dagger\B{b})(\B{b}^\dagger\B{b})^\tx{-1}(\B{b}^\dagger\B{y}) \notag \\
 &= \frac{1}{\sigma_n^2} \frac{ [\B{b}^\dagger(\BS\mu^\prime)\B{b}(\BS\mu)\gamma_{vv}^*] [\B{b}^\dagger(\BS\mu)\B{b}(\BS\mu^\prime)\gamma_{vv}] }{\B{b}^\dagger(\BS\mu)\B{b}(\BS\mu)}  \notag \\
 &= \frac{\vert \gamma_{vv} \vert^2}{\sigma_n^2} \frac{\vert \B{b}^\dagger(\BS\mu)\B{b}(\BS\mu^\prime)  \vert^2}{\lVert \B{b}(\BS\mu) \lVert^2}  \notag \\
 &\approx \rho_{vv} \frac{\vert \B{b}^\dagger(\BS\mu)\B{b}(\BS\mu^\prime)  \vert^2}{\lVert \B{b}(\BS\mu) \rVert \cdot \lVert \B{b}(\BS\mu^\prime) \rVert}  \notag\\
 &= \rho_{vv} \vert X_{tot} \vert ^2, \label{Eq:AmbCorr_Derive}
\end{align}
where the approximation assumes that $\lVert \B{b}(\BS\mu) \rVert$ has similar values for different $\BS\mu$. This is more or less fulfilled for antenna arrays such as \gls{uca} designed to cover signals from all possible azimuth angles. 

The derivation in Eq. (\ref{Eq:AmbCorr_Derive}) shows that our proposed ambiguity function is closely related with the correlation function and hence the \gls{mle} developed in Sect. \ref{Sect:HRPE}.

\section{Computation of 3D Tensors $\mathcal{T}_1$ and $\mathcal{T}_2$}
\label{Annex:InitialSPsTensor_RSAA}
Similar to the parameter initialization method in \cite[Appendix A]{WangHRPE2016}, we first construct a 4D tensor $\mathcal{Y}^\prime$ from $\B{y}$, then reorganize it by combining the data in the last two dimensions, i.e. \gls{tx} array and time, into one dimension. 
\begin{equation}
  \mathcal{Y}^\prime_\tx{3D} =\tx{reshape}(\mathcal{Y}^\prime, [M_f\; M_R\; M_TT]),
\end{equation}
where $\tx{reshape}()$ is a standard MATLAB function. The dimension of the 3D tensor $\mathcal{Y}^\prime_\tx{3D}$ is $M_f \times M_R \times M_TT$. As a result we can compute $\mathcal{T}_1^{n_t}$ as
\begin{equation}
  \mathcal{T}_1^{n_t} = \big[ (\mathcal{Y}^\prime_\tx{3D} \times_1 \B{B}_f^{\prime\dagger}) \times_2 \B{B}_{RV}^{n_t \prime \dagger} \big] \times_3 \B{B}_{TV,T}^{n_t \prime \dagger}.
\end{equation}
The expressions of $\B{B}_f^{\prime}$ and $\B{B}_{RV}^{n_t\prime}$ can be readily found in \cite[(61-62)]{WangHRPE2016}, while $\B{B}_{TV,T}^{n_t \prime}$ is a new component and given as follows.
\begin{align}
  \B{B}_{TV,T}^{n_t\prime} &= (\B{U}_t\otimes\B{U}_T)^\dagger \big(\B{B}_{TV,T} \odot \B{A}_{t,N_T}(\nu_{n_t}) \big),
\end{align}
where $\B{B}_{TV,T} \in \mathbb{C}^{M_TT \times N_T}$ is vertically stacked with $T$ copies of $\B{B}_{TV}$, and the matrix-valued function $\B{A}_{t,N_T}(\nu_{n_t})\in \mathbb{C}^{M_TT \times N_T}$ is constructed by horizontally stacking $N_T$ copies of the column vector $\B{a}_T(\nu_{n_t})$. This column vector is built by concatenating the columns of $\B{A}_T^t$ with $t=1,2,\ldots,T$, which are related with the $n_t$-th grid point of Doppler shift and given in (\ref{Eq:A_T_t}).


For another key element $\mathcal{T}_2^{n_t}$ used in Eq. (\ref{Eq:C_grid_4D}), we again can follow the method outlined in \cite[Appendix A]{WangHRPE2016} with some minor adjustments. First we need to build a 3D tensor $\mathcal{L}_i$ based on the eigenvalues of \B{R}, which is given by
\begin{equation}
  \mathcal{L}_i= \tx{reshape}\big( \tx{diag}(\BS\Lambda^\tx{-1},[M_f\;M_R\;M_TT])\big).
\end{equation}
Then the child-tensor $\mathcal{T}_2^{n_t}$ can be computed with tensor products according to
\begin{equation}
  \mathcal{T}_2^{n_t} = \big[ (\mathcal{L}_i \times_1 (\B{BB}_f^\prime)^\tx{T}) \times_2 (\B{BB}_{RV}^{n_t \prime})^\tx{T} \big] \times_3 (\B{BB}_{TV,T}^{n_t \prime})^\tx{T}.
\end{equation}

\section{Jacobian Matrix $\B{D}(\BS\theta_s)$}
\label{Append:JacobMat}
The Jacobian matrix $\B{D}(\BS\theta_s)$ is defined as
\begin{equation}
   \B{D}(\BS\theta_s) = \frac{\partial}{\partial \BS\theta_s^\tx{T}} \B{s}(\BS\theta_s),
\end{equation} 
with each column related to one partial derivative. More details about different columns in $\B{D}(\BS\theta_s)$ are given by
\begin{align}
 \frac{\partial}{\partial \BS\tau^\tx{T}}\B{s}(\BS\theta_\tx{sp}) &= \tilde{\B{B}}_{TV,T} \diamond \tilde{\B{B}}_{RV} \diamond \B{D}_f \diamond \BS\gamma_{vv}^\tx{T} \\
 \frac{\partial}{\partial \BS\varphi_T^\tx{T} } \B{s}(\BS\theta_\tx{sp}) &= \begin{bmatrix}
  \B{D}_{\varphi_T,V} \odot \B{A}_T^1 \\
  \vdots \\
  \B{D}_{\varphi_T,V} \odot \B{A}_T^\tx{T}
 \end{bmatrix} \diamond \tilde{\B{B}}_{RV} \diamond \B{B}_f \diamond \BS\gamma_{vv}^\tx{T} \\
 \frac{\partial}{\partial \BS\varphi_R^\tx{T} } \B{s}(\BS\theta_\tx{sp}) &=  \tilde{\B{B}}_{TV,T} \diamond (\B{D}_{\varphi_R,V}\odot \B{A}_R) \diamond \B{B}_f \diamond \BS\gamma_{vv}^\tx{T} \\
  \frac{\partial}{\partial \BS\gamma_{vv,r}^\tx{T}} \B{s}(\BS\theta_\tx{sp}) &=  \tilde{\B{B}}_{TV,T}  \diamond \tilde{\B{B}}_{RV} \diamond \B{B}_f  \\
 \frac{\partial}{\partial \BS\gamma_{vv,i}^\tx{T}} \B{s}(\BS\theta_\tx{sp}) &=  j \tilde{\B{B}}_{TV,T} \diamond \tilde{\B{B}}_{RV} \diamond \B{B}_f,
\end{align}
where $j$ stands for the unit imaginary number here. Particularly the partial derivative with respect to the Doppler shift is determined by
\begin{align}
 \frac{\partial}{\partial \BS\nu^\tx{T}} \B{s}(\BS\theta) &= \begin{bmatrix}
  \B{B}_{TV} \odot \B{D}_T^1 \\ \cdots \\ \B{B}_{TV} \odot \B{D}_T^t \\ \cdots \\ \B{B}_{TV} \odot \B{D}_T^\tx{T}
\end{bmatrix}  \diamond \tilde{\B{B}}_{RV} \diamond \B{B}_f \diamond \BS\gamma_{vv}^\tx{T} \notag \\
 &+  \tilde{\B{B}}_{TV,T} \diamond (\B{B}_{RV} \odot \B{D}_{\nu,R}) \diamond \B{B}_f \diamond \BS\gamma_{vv}^\tx{T}. 
\end{align}
Most of the \B{D} matrices in this appendix and Tab. \ref{Tab:DMatrix} are given in \cite[(78)-(83)]{WangHRPE2016} except for $\B{D}_T^j$ with $j=1,2,\ldots,T$, which is determined by $\B{D}_T^j = \mathcal{D}(\B{A}_T^j,\BS\nu)$. The operator $\mathcal{D}()$ is defined based on \cite[(77)]{WangHRPE2016}.

\bibliography{RSAA_draft_final}

\begin{thebibliography}{10}
\providecommand{\url}[1]{#1}
\csname url@samestyle\endcsname
\providecommand{\newblock}{\relax}
\providecommand{\bibinfo}[2]{#2}
\providecommand{\BIBentrySTDinterwordspacing}{\spaceskip=0pt\relax}
\providecommand{\BIBentryALTinterwordstretchfactor}{4}
\providecommand{\BIBentryALTinterwordspacing}{\spaceskip=\fontdimen2\font plus
\BIBentryALTinterwordstretchfactor\fontdimen3\font minus
  \fontdimen4\font\relax}
\providecommand{\BIBforeignlanguage}[2]{{%
\expandafter\ifx\csname l@#1\endcsname\relax
\typeout{** WARNING: IEEEtran.bst: No hyphenation pattern has been}%
\typeout{** loaded for the language `#1'. Using the pattern for}%
\typeout{** the default language instead.}%
\else
\language=\csname l@#1\endcsname
\fi
#2}}
\providecommand{\BIBdecl}{\relax}
\BIBdecl

\bibitem{molisch2012wireless}
A.~F. Molisch, \emph{Wireless Communications}, 2nd~ed.\hskip 1em plus 0.5em
  minus 0.4em\relax IEEE Press - Wiley, 2011.

\bibitem{HurMagzine}
S.~Hur and et~al., ``Feasibility of mobility for millimeter-wave systems based
  on channel measurement,'' \emph{IEEE Communications Magazine}, submitted.

\bibitem{mecklenbrauker2011vehicular}
C.~F. Mecklenbrauker, A.~F. Molisch, J.~Karedal, F.~Tufvesson, A.~Paier,
  L.~Bernad{\'o}, T.~Zemen, O.~Klemp, and N.~Czink, ``Vehicular channel
  characterization and its implications for wireless system design and
  performance,'' \emph{Proceedings of the IEEE}, vol.~99, no.~7, pp.
  1189--1212, 2011.

\bibitem{wang2016channel}
C.-X. Wang, A.~Ghazal, B.~Ai, Y.~Liu, and P.~Fan, ``Channel measurements and
  models for high-speed train communication systems: a survey,'' \emph{IEEE
  communications surveys \& tutorials}, vol.~18, no.~2, pp. 974--987, 2016.

\bibitem{kim2015large}
M.~Kim, J.-i. Takada, Y.~Chang, J.~Shen, and Y.~Oda, ``Large scale
  characteristics of urban cellular wideband channels at 11 {GHz},'' in
  \emph{Antennas and Propagation (EuCAP), 2015 9th European Conference
  on}.\hskip 1em plus 0.5em minus 0.4em\relax IEEE, 2015, pp. 1--4.

\bibitem{martin1998spatio}
U.~Martin, ``Spatio-temporal radio channel characteristics in urban
  macrocells,'' \emph{IEE Proceedings-Radar, Sonar and Navigation}, vol. 145,
  no.~1, pp. 42--49, 1998.

\bibitem{thoma2000identification}
R.~S. Thoma, D.~Hampicke, A.~Richter, G.~Sommerkorn, A.~Schneider,
  U.~Trautwein, and W.~Wirnitzer, ``Identification of time-variant directional
  mobile radio channels,'' \emph{IEEE Transactions on Instrumentation and
  measurement}, vol.~49, no.~2, pp. 357--364, 2000.

\bibitem{yin2003doppler}
X.~Yin, B.~H. Fleury, P.~Jourdan, and A.~Stucki, ``Doppler frequency estimation
  for channel sounding using switched multiple-element transmit and receive
  antennas,'' in \emph{Global Telecommunications Conference, 2003. GLOBECOM'03.
  IEEE}, vol.~4.\hskip 1em plus 0.5em minus 0.4em\relax IEEE, 2003, pp.
  2177--2181.

\bibitem{fleury2003high}
B.~H. Fleury, X.~Yin, P.~Jourdan, and A.~Stucki, ``High-resolution channel
  parameter estimation for communication systems equipped with antenna
  arrays,'' in \emph{Proc. 13th Ifac Symposium on System Identification (sysid
  2003)}, 2003.

\bibitem{pedersen2004joint}
T.~Pedersen, C.~Pedersen, X.~Yin, B.~H. Fleury, R.~R. Pedersen, B.~Bozinovska,
  A.~Hviid, P.~Jourdan, and A.~Stucki, ``Joint estimation of {Doppler}
  frequency and directions in channel sounding using switched {Tx} and {Rx}
  arrays,'' in \emph{Global Telecommunications Conference, 2004. GLOBECOM'04.
  IEEE}, vol.~4.\hskip 1em plus 0.5em minus 0.4em\relax IEEE, 2004, pp.
  2354--2360.

\bibitem{pedersen2008optimization}
T.~Pedersen, C.~Pedersen, X.~Yin, and B.~H. Fleury, ``Optimization of
  spatiotemporal apertures in channel sounding,'' \emph{Signal Processing, IEEE
  Transactions on}, vol.~56, no.~10, pp. 4810--4824, 2008.

\bibitem{belloni2007doa}
F.~Belloni, A.~Richter, and V.~Koivunen, ``Doa estimation via manifold
  separation for arbitrary array structures,'' \emph{IEEE Transactions on
  Signal Processing}, vol.~55, no.~10, pp. 4800--4810, 2007.

\bibitem{eric1998ambiguity}
M.~Eri{\'c}, A.~Zejak, and M.~Obradovi{\'c}, ``Ambiguity characterization of
  arbitrary antenna array: Type {I} ambiguity,'' in \emph{Spread Spectrum
  Techniques and Applications, 1998. Proceedings., 1998 IEEE 5th International
  Symposium on}, vol.~2.\hskip 1em plus 0.5em minus 0.4em\relax IEEE, 1998, pp.
  399--403.

\bibitem{chen2008mimo}
C.-Y. Chen and P.~Vaidyanathan, ``{MIMO} radar ambiguity properties and
  optimization using frequency-hopping waveforms,'' \emph{Signal Processing,
  IEEE Transactions on}, vol.~56, no.~12, pp. 5926--5936, 2008.

\bibitem{wang2018sw}
R.~Wang, O.~Renaudin, C.~U. Bas, S.~Sangodoyin, and A.~F. Molisch, ``Antenna
  switching sequence design for channel sounding in a fast time-varying
  channel,'' in \emph{Communications (ICC), IEEE International Conference
  on}.\hskip 1em plus 0.5em minus 0.4em\relax IEEE, 2018 (accepted).

\bibitem{salmi2009detection}
J.~Salmi, A.~Richter, and V.~Koivunen, ``Detection and tracking of {MIMO}
  propagation path parameters using state-space approach,'' \emph{Signal
  Processing, IEEE Transactions on}, vol.~57, no.~4, pp. 1538--1550, 2009.

\bibitem{Wang2015efficiency}
R.~Wang, O.~Renaudin, R.~M. Bernas, and A.~F. Molisch, ``Efficiency improvement
  for path detection and tracking algorithm in a time-varying channel,'' in
  \emph{Vehicular Technology Conference (VTC fall), 2015 IEEE 82nd}.\hskip 1em
  plus 0.5em minus 0.4em\relax IEEE, 2015.

\bibitem{WangHRPE2016}
R.~Wang, O.~Renaudin, C.~U. Bas, S.~Sangodoyin, and A.~F. Molisch,
  ``High-resolution parameter estimation for time-varying double directional
  {V2V} channel,'' \emph{IEEE Transactions on Wireless Communications},
  vol.~16, no.~11, pp. 7264--7275, 2017.

\bibitem{kay1993fundamentals}
S.~M. Kay, ``Fundamentals of statistical signal processing, volume i:
  Estimation theory (v. 1),'' \emph{PTR Prentice-Hall, Englewood Cliffs}, 1993.

\bibitem{wang2017real}
R.~Wang, C.~U. Bas, O.~Renaudin, S.~Sangodoyin, U.~T. Virk, and A.~F. Molisch,
  ``A real-time {MIMO} channel sounder for vehicle-to-vehicle propagation
  channel at 5.9 {GHz},'' in \emph{Communications (ICC), 2017 IEEE
  International Conference on}.\hskip 1em plus 0.5em minus 0.4em\relax IEEE,
  2017, pp. 1--6.

\bibitem{kirkpatrick1983optimization}
S.~Kirkpatrick, C.~D. Gelatt, M.~P. Vecchi \emph{et~al.}, ``Optimization by
  simulated annealing,'' \emph{science}, vol. 220, no. 4598, pp. 671--680,
  1983.

\bibitem{richter2005estimation}
A.~Richter, ``Estimation of radio channel parameters: Models and algorithms,''
  Ph.D. dissertation, Techn. Univ. Ilmenau, Ilmenau, Germany, May 2005.
  [Online]. Available: http://www.db-thueringen.de.

\bibitem{TTB_Software}
\BIBentryALTinterwordspacing
B.~W. Bader, T.~G. Kolda \emph{et~al.}, ``{MATLAB} tensor toolbox version
  2.5,'' Available online, January 2012. [Online]. Available:
  \url{http://www.sandia.gov/~tgkolda/TensorToolbox/}
\BIBentrySTDinterwordspacing

\bibitem{abbas2011directional}
T.~Abbas, J.~Karedal, F.~Tufvesson, A.~Paier, L.~Bernad{\'o}, and A.~F.
  Molisch, ``Directional analysis of vehicle-to-vehicle propagation channels,''
  in \emph{Vehicular Technology Conference (VTC Spring), 2011 IEEE 73rd}.\hskip
  1em plus 0.5em minus 0.4em\relax IEEE, 2011, pp. 1--5.

\bibitem{wang2017pimrc}
R.~Wang, O.~Renaudin, C.~U. Bas, S.~Sangodoyin, and A.~F. Molisch,
  ``Vehicle-to-vehicle propagation channel for truck-to-truck and mixed
  passenger freight convoy,'' in \emph{Personal, Indoor, and Mobile Radio
  Communications (PIMRC), 2017 IEEE 28th Annual International Symposium
  on}.\hskip 1em plus 0.5em minus 0.4em\relax IEEE, 2017, pp. 1--5.

\end{thebibliography}
\bibliographystyle{IEEEtran}

\end{document}